\documentclass[preprint,3p,10pt]{elsarticle}

\usepackage{amssymb}

\newdefinition{rmk}{Remark}

\usepackage{algorithm}
\usepackage[end]{algpseudocode}
\usepackage{amsmath}
\usepackage{booktabs}
\usepackage{caption}
\usepackage{setspace}
\usepackage{siunitx}
\usepackage{subcaption}
\usepackage{xcolor}
\usepackage{hyperref}

\usepackage{tikz}
\usepackage{standalone}
\usetikzlibrary{arrows.meta}
\usetikzlibrary{shapes.misc}
\usetikzlibrary{shapes.geometric}
\usetikzlibrary{calc}

\biboptions{sort&compress}

\renewcommand{\vec}[1]{\boldsymbol{\mathbf{#1}}}         
\newcommand{\mat}[1]{\boldsymbol{\mathbf{#1}}}           
\newcommand{\norm}[1]{\| #1 \|}                 
\newcommand{\abs}[1]{| #1 |}                    

\journal{arXiv}

\begin{document}

\begin{frontmatter}


\title{Towards Additively Manufactured Metamaterials with Powder Inclusions for Controllable Dissipation: The Critical Influence of Packing Density}

\author[LNM]{Patrick M. Praegla\corref{cor}}
\ead{patrick.praegla@tum.de}

\author[IWB]{Thomas Mair}
\ead{thomas.mair@iwb.tum.de}

\author[IWB]{Andreas Wimmer}
\ead{andreas.wimmer@iwb.tum.de}

\author[LNM]{Sebastian L. Fuchs}
\ead{sebastian.fuchs@tum.de}

\author[LNM]{Niklas Fehn}
\ead{niklas.fehn@tum.de}

\author[IWB]{Michael F. Zaeh}
\ead{michael.zaeh@iwb.tum.de}

\author[LNM]{Wolfgang A. Wall}
\ead{wolfgang.a.wall@tum.de}

\author[LNM]{Christoph Meier}
\ead{christoph.anton.meier@tum.de}

\cortext[cor]{corresponding author}

\affiliation[LNM]{organization={Institute of Computational Mechanics, Technical University of Munich},
            addressline={Boltzmannstra{\ss}e 15}, 
            city={Garching b. M\"unchen},
            postcode={85748}, 
            state={Bavaria},
            country={Germany}}

\affiliation[IWB]{organization={Institute for Machine Tools and Industrial Management, Technical University Munich},
            addressline={Boltzmannstra{\ss}e 15}, 
            city={Garching b. M\"unchen},
            postcode={85748}, 
            state={Bavaria},
            country={Germany}}

\begin{abstract}
Particle dampers represent a simple yet effective means to reduce unwanted oscillations when attached to structural components. Powder bed fusion additive manufacturing of metals allows to integrate particle inclusions of arbitrary shape, size and spatial distribution directly into bulk material, giving rise to novel metamaterials with controllable dissipation without the need for additional external damping devices. At present, however, it is not well understood how the degree of dissipation is influenced by the properties of the enclosed powder packing. In the present work, a two-way coupled discrete element - finite element model is proposed allowing for the first time to consistently describe the interaction between oscillating deformable structures and enclosed powder packings. As fundamental test case, the free oscillations of a hollow cantilever beam filled with various powder packings differing in packing density, particle size, and surface properties are considered to systematically study these factors of influence. Critically, it is found that the damping characteristics strongly depend on the packing density of the enclosed powder and that an optimal packing density exists at which the dissipation is maximized. Moreover, it is found that the influence of (absolute) particle size on dissipation is rather small. First-order analytical models for different deformation modes of such powder cavities are derived to shed light on this observation.
\end{abstract}




\end{frontmatter}


\section{Introduction}
\label{sec:introduction}
Particle dampers represent a simple yet effective means to reduce unwanted oscillations when attached to structural components. While conventional production technologies require to realize particle dampers as additional external devices, Powder Bed Fusion of Metals Using a Laser Beam (PBF-LB/M) allows to integrate particle inclusions of arbitrary shape, size, and spatial distribution directly into bulk material, giving rise to novel meta materials with controllable dissipation~\cite{Ehlers2021a,Scott-Emuakpor2021,Kuenneke2017,Vogel2019}. Importantly, PBF-LB/M naturally allows to use the powder particles underlying the process as raw material also as means of dissipation. By creating closed cavities, pockets of unfused powder remain inside the part after the process. At present, however, it is not well understood how the degree of dissipation is influenced by the properties of the enclosed powder packing. In the present work, a two-way coupled discrete element - finite element model is proposed allowing for the first time to consistently describe the interaction between oscillating structures and enclosed powder packings and to systematically study the influence of powder packing characteristics on the resulting degree of dissipation.

Particle dampers have been studied to great extent in the literature with applications in civil and mechanical engineering. Compared to other damping mechanisms, they provide various advantages. Being a passive damping mechanism, they do not require an additional energy source. They are simple in design, easy to install at different locations, work in multiple directions and over a wide frequency range, and are insensitive to ambient temperature~\cite{Gagnon2019}. Though, in additive manufacturing high temperatures from post heat treatment should be avoided as the unfused particles may sinter and thus significantly decrease the damping capabilities~\cite{Kuenneke2017,Scott-Emuakpor2021, Ehlers2022}. In experimental studies particle dampers are most often attached as external devices. In numerical and analytical studies of such particle dampers, the cantilever beam is often simplified to a single degree of freedom oscillator to reduce the model complexity~\cite{Mao2004,Lu2011, Lu2018,Masmoudi2016,Gagnon2019}. Particle dampers are typically characterized by a highly non-linear behavior that depends on various parameters such as particle size, mass, coefficient of friction, and coefficient of restitution~\cite{Lu2011}. Comprehensive overviews of conventional particle dampers can be found in the review articles by Lu et al.~\cite{Lu2018} and Gagnon et al.~\cite{Gagnon2019}. Here and in the following, the notion conventional particle dampers refers to devices produced by conventional manufacturing technologies. These systems commonly consist of an enclosure that is partially filled with rather large particles in the millimeter range and that is externally attached to the structure to be damped.

Additively manufactured particle dampers need to be studied separately as they differ from conventional particle dampers in various aspects such that a transferability of results is questionable. In particular, the particle sizes typically differ, lying in the millimeter range for conventional particle dampers (cf. Table 3 in~\cite{Gagnon2019}) and in the micrometer range in PBF-LB/M, e.g., with a mean diameter of~\SI{27}{\micro\meter} in~\cite{Scott-Emuakpor2021} and~\SI{47}{\micro\meter} in~\cite{Ehlers2021}. Another key feature is the filling level, or packing density, within a particle damper. When manufacturing closed cavities with PBF-LB/M, the amount of powder particles inside the cavity is defined by the process parameters of the PBF-LB/M and the previous powder spreading process. These cavities are completely filled with powder typically characterized by packing densities in the range of~\SIrange{40}{60}{\percent}~\cite{Kuenneke2017,Ehlers2021,Scott-Emuakpor2019}. For conventional particle dampers, it was found that a certain clearance between the particles and the walls of the damper is necessary to increase the degree of damping, resulting in an effective packing density that is rather in the range of~\SI{10}{\percent}~\cite{Mao2004}. This conceptual difference is likely to change the fundamental mechanism of dissipation. While dynamic particle impacts can be considered as dominating means of dissipation in conventional particle dampers, stick-slip effects between particles in permanent contact might be the main source of dissipation in additively manufactured particle dampers.

With these differences in mind, the literature specifically concerned with additively manufactured particle dampers is summarized in the following. In experiments, K{\"u}nneke and Zimmer~\cite{Kuenneke2017} attached cuboid cavities filled with powder to a spring element and measured the free decay from an initial excitation to study the influence of different cavity geometries. They found that a larger cavity provided better damping. Further, a subdivision of the cavity or lattice structures inside the cavity decreased the amount of damping. In a similar way, Guo et al.~\cite{Guo2021} studied a multi-unit particle damper by attaching the additively manufactured part to the end of a cantilever beam, adopting the same experimental setup as for a conventional particle damper~\cite{Friend2000,Marhadi2005}. For the studied low vibration frequency ($< \SI{100}{\hertz}$), one large unit cell yielded better damping performance than multiple smaller cells. Additionally, a DEM model was used to study the energy dissipation. The computational effort was reduced by considering only one unit cell of size~$2\times 2 \times 2~\si{\milli\meter\cubed}$ and using mono-sized particles of~\SI{30}{\micro\meter} diameter while the real particle size ranges from~\SIrange{15}{40}{\micro\meter}. The damping mechanism was found to be mainly caused by impacts of the particles with the walls.
For their experiments, Ehlers et al.~\cite{Ehlers2021} manufactured complete beams of AlSi10Mg with closed cavities containing unfused powder. The beams were excited by an automatic impulse hammer. The best damping was achieved for lower natural frequencies while damping decreased with increasing natural frequency. This correlation between damping ratio and natural frequency was approximated with a hyperbola. Further, a $3^3$ full factorial experimental design was developed in~\cite{Ehlers2022} to identify the influence of excitation force, excitation frequency, and cavity size for the aluminum alloy AlSi10Mg and the tool steel~1.2709. The amplitude of the frequency response function was reduced by up to~\SI{97}{\percent} due to the particle damper.
Similarly, Scott-Emuakpor et al.~\cite{Scott-Emuakpor2018,Scott-Emuakpor2019,Scott-Emuakpor2020,Scott-Emuakpor2021,Scott-Emuakpor2021a,Scott-Emuakpor2021b} published several studies on additively manufactured cantilever beams with multiple cavities excited by a shaker. Up to ten times the damping of a fully-fused beam could be reached with only~\SIrange{1}{4}{\percent} unfused powder volume~\cite{Scott-Emuakpor2018}. Using the experimental data from~\cite{Scott-Emuakpor2018,Scott-Emuakpor2019,Scott-Emuakpor2020,Scott-Emuakpor2021,Scott-Emuakpor2021a,Scott-Emuakpor2021b}, Kiracofe et al.~\cite{Kiracofe2021} created a discrete element method (DEM) model to predict the damping ratio of particle dampers. Since the entire cavity with 26 million particles at an average diameter of~\SI{25}{\micro\meter} was not feasible to simulate, the domain was subdivided into 5000 identical subdomains with 5000 particles each such that only one subdomain needed to be simulated. This subdomain was attached to a single degree of freedom spring-damper system excited by an external force. The damping performance was calculated from the response to sine dwell excitations near resonance. The parameters of the model were chosen to match the experimental data best. Using five validation points they predicted similar trends as in the experiments. Harduf et al.~\cite{Harduf2023} recently proposed a two-mass model for the steady-state response of structures containing particle dampers. The model consisted of a spring-dashpot system where the loose powder particles were modeled as an additional lumped mass which was coupled to the system via Coulomb friction. 

In summary, only a few works considered the modeling and numerical simulation of additively manufactured particle dampers. Guo et al.~\cite{Guo2021} used only mono-sized particles which can lower the damping performance~\cite{Kiracofe2021}. Further, the initial face-centered cubic particle configuration yielded a rather high packing density of~\SI{72}{\percent} while a typical packing density of spread powder is in the range of~\SIrange{40}{60}{\percent}~\cite{Meier2019a}. Kiracofe et al.~\cite{Kiracofe2021} only simulated a subdomain of the entire powder cavity. Both publications reduced the oscillating structure to a single degree of freedom oscillator with an attached rigid box, however, deformation of the powder cavity was not considered. Such models considering rigid particle boxes allow to study dissipation due to powder particle impacts, i.e. energy from the enclosure is transferred to the powder only via impacts. However, to accurately model frictional dissipation due to tangential relative motion of powder particles, which is assumed to be the main source of dissipation for densely packed powder cavities, consideration of a deformable enclosure is imperative. 

To close this gap left by existing approaches, the present work proposes a two-way coupled discrete element~-~finite element model to consistently describe the interaction between oscillating deformable structures and powder packings enclosed in cavities within these structures. Importantly, this approach allows for the first time to consider the relative slip motion induced on the powder particles through the deformation of the cavity walls, which is believed to be a major source of dissipation when considering completely filled and densely packed powder cavities embedded in deformable structures, as typically resulting from PBF-LB/M processes. The powder domain (represented by the DEM) and the structural domain (discretized by the finite element method (FEM)) are coupled via a Dirichlet-Neumann partitioned approach. As fundamental test case, the free oscillations of a hollow cantilever beam filled with powder packings differing in packing density as well as particle size, density and surface properties are considered to study these factors of influence systematically. Critically, it is found that the damping characteristics strongly depend on the packing density of the enclosed powder and that an optimal packing density exists at which the dissipation is maximized. Moreover, it is found that the influence of (absolute) particle size on dissipation is rather small. First-order analytical models for different deformation modes of such powder cavities are derived to shed light on this observation.

The remainder of this article is organized as follows. First, the DEM powder model, the FEM discretization of the solid domain as well as the DEM-FEM coupling approach are summarized, followed by a description of the simulation setup. Next, the computational model is used to study the influence of different parameters of the model on the dissipation behavior of particle dampers by means of cantilever beams. An analytical model is derived to study the influence of the particle size. Finally, an experimental realization of the simulation setup is presented.

\section{Methods}
\label{sec:methods}
To enable numerical studies of particle dampers, the discrete element method (DEM) is employed to model the powder particles within the damper. Additionally, the DEM is coupled with the finite element method (FEM) which is used to discretize the (deformable) structural enclosure of the particle damper. The domain $\Omega$ of the particle-structure interaction problem consists of a non-overlapping domain $\Omega^p$ filled with particles and a structural domain $\Omega^s$ that share a common interface $\Gamma^{ps}$.
\subsection{DEM powder model}
\label{sec:dem}
To model the powder phase, a cohesive powder model recently proposed by the authors in the context of powder spreading in powder bed fusion processes \cite{Meier2019,Meier2019a,Meier2021,Penny2021}, will be employed. In the following, only the most important equations are summarized.
Using the DEM, the following equations of motion are solved for each particle $i$ in each time step:
\begin{equation}
    \label{eq:governing_equations}
    \begin{split}
        & (m\ddot{\vec{r}}_G)^i = m^i\vec{g} + \sum_j (\vec{f}_{CN}^{ij} + \vec{f}_{CT}^{ij} + \vec{f}_{AN}^{ij}) \quad \text{in} \quad \Omega^p,\\
        & (I_G\dot{\vec{\omega}})^i = \sum_j (\vec{m}_{R}^{ij} + \vec{r}_{CG}^{ij} \times \vec{f}_{CT}^{ij}) \quad \text{in} \quad \Omega^p,
    \end{split}
\end{equation}
with the mass $m=\frac{4}{3}\pi\rho r^3$ and moment of inertia $I_G = 0.4mr^2$ for spherical particles with radius $r$ and density $\rho$. The interaction forces consist of normal contact $\vec{f}_{CN}^{ij}$, frictional contact $\vec{f}_{CT}^{ij}$, and adhesive $\vec{f}_{AN}^{ij}$ forces. In the balance of angular momentum, moment contributions from the frictional forces with lever arm $\vec{r}_{CG}^{ij} = \vec{r}_{C}^{ij} - \vec{r}_{G}^{i}$ from the particle center to the point of contact, and from the rolling resistance $\vec{m}_{R}^{ij}$ are considered. Each contribution is briefly explained in the following. Normal contact forces are evaluated with a spring-dashpot model
\begin{equation}
	\vec{f}_{CN}^{ij} = \begin{cases}
		\min(0,k_N g_N + d_N \dot{g}_N)\vec{n}, &g_N \le 0, \\
		\vec{0}, &g_N > 0,
	\end{cases}
\end{equation}
where $g_N$ is the normal gap (penetration) and $\vec{n}$ the normal vector between two particles according to
\begin{equation}
	g_N := \norm{\vec{r}_G^j - \vec{r}_G^i} - (r_i + r_j),
	\qquad
	\vec{n} = \frac{\vec{r}_G^j - \vec{r}_G^i}{\norm{\vec{r}_G^j - \vec{r}_G^i}}.
\end{equation}
The stiffness constant $k_N$ and the damping constant $d_N$ are given by
\begin{equation}
	k_N \ge \max\left(\frac{8\pi\rho V_{max}^2 r_{max}}{c_g^2}, \frac{4\pi\gamma}{c_g}\right),
	\quad
	d_N = 2 \abs{\ln(e)} \sqrt{\frac{k_N m_{eff}}{\ln(e)^2+\pi^2}},
\end{equation}
where the contact stiffness $k_N$ is chosen such that dynamic collisions of particles with the maximum radius $r_{max}$ and maximum velocity $V_{max}$ or the maximum static adhesive forces, characterized by the surface energy $\gamma$, lead to penetrations limited by the maximum penetration $c_g$. The damping constant is characterized by the coefficient of restitution $e$ and the effective mass $m_{eff} = \frac{m_i m_j}{m_i + m_j}$.

Similarly, the frictional forces follow a spring-dashpot model coupled to the normal force via Coulomb's law
\begin{equation} \label{eq:frictional_forces}
	\vec{f}_{CT}^{ij} = \begin{cases}
		\min (\mu \norm{\vec{f}_{CN}^{ij}}, \norm{k_T\vec{g}_T + d_T \dot{\vec{g}}_T})\vec{t}_T, & g_N \le 0, \\
		\vec{0}, & g_N > 0,
	\end{cases}
\end{equation}
with the coefficient of friction $\mu$, the constants $k_T = \frac{1-\nu}{1-0.5\nu}k_N$ and $d_T = d_N$, where $\nu$ is Poisson's ratio, and the tangential gap vector $\vec{g}_T$, its rate $\dot{\vec{g}}_T$ and the tangential unit vector $\vec{t}_T$.

The rolling resistance follows a spring-dashpot model as well. Though, simulations showed that the influence of the rolling resistance is negligible for the considered particle dampers and is therefore left out here.

The adhesive forces result from van-der-Waals forces between particles and are characterized by the pull-off force, i.e., the force necessary to separate two contacting particles. The pull-off force is calculated from the effective surface energy via the Derjaguin-Muller-Toporov (DMT) model~\cite{Derjaguin1975}. The surface energy, used to characterize the magnitude of adhesive forces, is greatly affected by the particle properties (e.g., the roughness and surface contaminations) and is therefore usually calibrated with experiments. As later shown, adhesive forces only play a minor role for the dissipation in the considered particle damper system. For time integration a velocity-verlet scheme is used. More details on the overall powder model can be found in~\cite{Meier2019}.

\subsection{FEM discretization of solid domain}
\label{sec:fem}
Considering the regime of finite deformations, the structural field is governed by the nonlinear balance of linear momentum in the following local material form:
\begin{equation} \label{eq:balance_momentum}
\rho^{s}_{0} \frac{\mathrm{d}^2 \mathbf{d}^{s}}{\mathrm{d}t^2} = \nabla_{0} \cdot \left(\vec{F} \vec{S}\right) + \rho_{0}^{s} \vec{b}_{0}^{s} \quad \text{in} \quad \Omega^{s},
\end{equation}
with the reference density~$\rho^{s}_{0}$ and body force~$\vec{b}_{0}^{s}$, and the structural displacement~$\vec{d}^{s}$ as primary unknowns. The deformation of the structure is described by the deformation gradient $\vec{F} = \nabla_{0} {\vec{d}^{s}}$ defining the Green-Lagrange strains $\mat{E} = \frac{1}{2} \left(\mat{F}^{T} \mat{F} - \mat{I}\right)$. For simplicity, the second Piola-Kirchhoff stresses $\mat{S}$ are chosen to follow from a constitutive relation of the form $\mat{S} = \partial\Psi/\partial\mat{E}$ based on a hyperelastic strain energy function $\Psi = \Psi\left(\mat{E}\right)$. Within this work, a Saint Vernant Kirchoff constitutive law is chosen. The partial differential equation~\eqref{eq:balance_momentum} is subject to initial conditions for the structural displacement and velocity field:
\begin{equation}
\vec{d}^{s} = \vec{d}_{0}^{s} \quad \text{and} \quad \frac{\mathrm{d}\vec{d}^{s}}{\mathrm{d}t} = \frac{\mathrm{d}\vec{d}_{0}^{s}}{\mathrm{d}t} \quad \text{in} \quad \Omega^{s} \quad \text{at} \quad t = 0 \, .
\end{equation}
On the structural boundary~$\Gamma^{s} = \partial\Omega^{s} \setminus \Gamma^{ps}$, where $\partial\Omega^{s}$ is the total boundary of the structural domain and $\Gamma^{ps}$ the powder structure interface, Dirichlet and Neumann boundary conditions are prescribed
\begin{equation}
\vec{d}^{s} = \vec{\hat{d}}^{s} \quad \text{on} \quad \Gamma_{D}^{s} \quad \text{and} \quad 
\left(\mat{F} \mat{S}\right) \cdot \vec{N} = \vec{\hat{t}}_{0}^{s}  \quad \text{on} \quad \Gamma^{s}_{N} \, ,
\end{equation}
with prescribed boundary displacement~$\vec{\hat{d}}^{s}$, (first Piola-Kirchoff) boundary traction~$\vec{\hat{t}}^{s}_{0}$, and outward pointing unit normal vector~$\vec{N}$ on~$\Gamma^{s}$ in material description, where $\Gamma^{s} = \Gamma_{D}^{s} \cup \Gamma_{N}^{s}$ and $\Gamma_{D}^{s} \cap \Gamma_{N}^{s} = \emptyset$.

The balance of linear momentum~\eqref{eq:balance_momentum} is discretized by the finite element method. For more details on this standard procedure, the interested reader is referred to the literature~\cite{Hughes2012, Zienkiewicz2014}. The general form of the resulting semi-discrete equations of motion is given by
\begin{equation} \label{eq:fem:semi-discrete}
    \mat{M}\ddot{\vec{d}} + \vec{F}_{int}(\vec{d}) - \vec{F}_{ext} = 0,
\end{equation}
with the mass matrix $\mat{M}$, the vector of nonlinear internal forces $\vec{F}_{int}$, the vector of external forces $\vec{F}_{ext}$ and the time-dependent displacement vector $\vec{d}$. The semi-discrete problem~\eqref{eq:fem:semi-discrete} is discretized in time with the \mbox{generalized-$\alpha$} method~\cite{Chung1993} choosing a spectral radius of $\rho_{\infty} = 0.8$.

\begin{rmk}
    Damping for the structural part ($\mat{C}\dot{\vec{d}}$) is not considered. Though, it could be easily incorporated, e.g., with Rayleigh damping. Neglecting the structural damping allows to isolate the effect of the particle damper and attribute the dissipation to the particle interactions.
\end{rmk}

\subsection{DEM-FEM coupling approach}
\label{sec:coupling}
The powder and structural domains are coupled via a Dirichlet-Neumann partitioned approach, similar to the authors' previous work~\cite{Fuchs2021}, where a fluid and structural problem were discretized by smoothed particle hydrodynamics (SPH) and the FEM, respectively. The DEM particles are the Dirichlet partition with prescribed interface displacements $\vec{d}^{ps}$ at the particle-structure interface~$\Gamma^{ps}$. The structural field is the Neumann partition subject to the forces $\vec{f}^{ps}$ transfered by the particles.

Introducing the field operators~$\mathcal{P}$ and~$\mathcal{S}$ for the particle and structural problem, both mapping the interface displacements~$\vec{d}^{ps}$ to interface forces
\begin{equation}
\vec{f}^{ps}_{\mathcal{P}} = \mathcal{P}(\vec{d}^{ps}) \quad \text{and} \quad \vec{f}^{ps}_{\mathcal{S}} = \mathcal{S}(\vec{d}^{ps}) \, ,
\end{equation}
equilibrium at the interface~$\Gamma^{ps}$ is satisfied in case the condition
\begin{equation}
\mathcal{P}(\vec{d}^{ps}) = \mathcal{S}(\vec{d}^{ps})
\end{equation}
holds. The inverse particle and structural field operators mapping interface forces~$\vec{f}^{ps}$ to interface displacements are consequently defined as
\begin{equation}
\vec{d}^{ps}_{\mathcal{P}} = \mathcal{P}^{-1}(\vec{f}^{ps}) \quad \text{and} \quad \vec{d}^{ps}_{\mathcal{S}} = \mathcal{S}^{-1}(\vec{f}^{ps}) \, .
\end{equation}

Contact points of particles with surface elements are determined via closest point projection. The resulting interface forces acting at the respective closest points $\boldsymbol{\xi}$ are distributed to the nodes $j$ of interface element $e$ using its shape functions $N_j^e$ evaluated at the closest point. The nodal interface force $\vec{f}_{j}^{ps}$ at node $j$ results from the summation of all force contributions $\vec{f}_{i}^{e}$ of particles $i$ acting on various interface elements $e$ adjacent to node $j$:
\begin{equation}
    \vec{f}_{j}^{ps} = \sum_e \sum_i N_j^e(\boldsymbol{\xi}_i) \vec{f}_{i}^{e} \, .
\end{equation}

The particle and structural problem are solved repeatedly using an iterative fixed-point solver~\cite{Kuettler2008} until the convergence criterion  
\begin{equation} \label{eq:conv_crit}
    \frac{\left|\Delta{}\vec{d}^{ps}_{n+1,k+1}\right|}{\Delta{}t \sqrt{n^{ps}_{dof}}} < \epsilon
\end{equation}
is satisfied, with the L${}_2$-norm of the increment of interface displacements $\left|\Delta\vec{d}^{ps}_{n+1,k+1}\right| = \left|\vec{d}_{n+1}^{ps} - \vec{d}_{n}^{ps}\right|$ at time step $n$ and iteration $k$, the time step size $\Delta{}t$, the number of degrees of freedom at the interface $n_{dof}^{ps}$, and the tolerance $\epsilon$ (set to $\epsilon = 1\cdot 10^{-3}$ in the present work). The coupling algorithm is explained in detail in Algorithm~\ref{alg:part_pasi}.

\begin{algorithm}[htbp]
\caption{Time loop until final time $T$ of a Dirichlet-Neumann partitioned fixed-point particle-structure interaction algorithm} \label{alg:part_pasi}
\begin{algorithmic}[0]\onehalfspacing
\While{$t < T$}
        \State $t \gets t + \Delta{t}$
        \Comment increment time
        \State $k \gets 1$
        \Comment reset iteration counter
        \State $\vec{d}^{ps}_{n+1,k}$
        \Comment predict interface displacements
        \While{$true$}
                \State $\vec{f}^{ps}_{n+1,k+1} = \mathcal{P}\left(\vec{d}^{ps}_{n+1,k}\right)$
                \Comment solve particle field
                \State $\vec{d}^{ps}_{n+1,k+1} = \mathcal{S}^{-1}\left(\vec{f}^{ps}_{n+1,k+1}\right)$
                \Comment solve structural field
                \State $\Delta\vec{d}^{ps}_{n+1,k+1} = \vec{d}^{ps}_{n+1,k+1} - \vec{d}^{ps}_{n+1,k}$
                \Comment compute increment of interface displacements
                \If {$ \big|\Delta\vec{d}^{ps}_{n+1,k+1}\big| \bigg/ \left(\Delta{}t \, \sqrt{n^{ps}_{dof}}\right) < \epsilon$}
                \Comment check convergence criterion, cf. equation~\eqref{eq:conv_crit}
                        \State \textbf{break}
                \EndIf
                \State $k \gets k + 1$
                \Comment increment iteration counter
        \EndWhile
        \State $n \gets n + 1$
        \Comment increment step counter
\EndWhile
\end{algorithmic}
\end{algorithm}

\subsection{Simulation setup}
\label{sec:sim_setup}
\begin{figure}
    \centering
    \begin{subfigure}[t]{0.24\textwidth}
        \includegraphics[width=\textwidth]{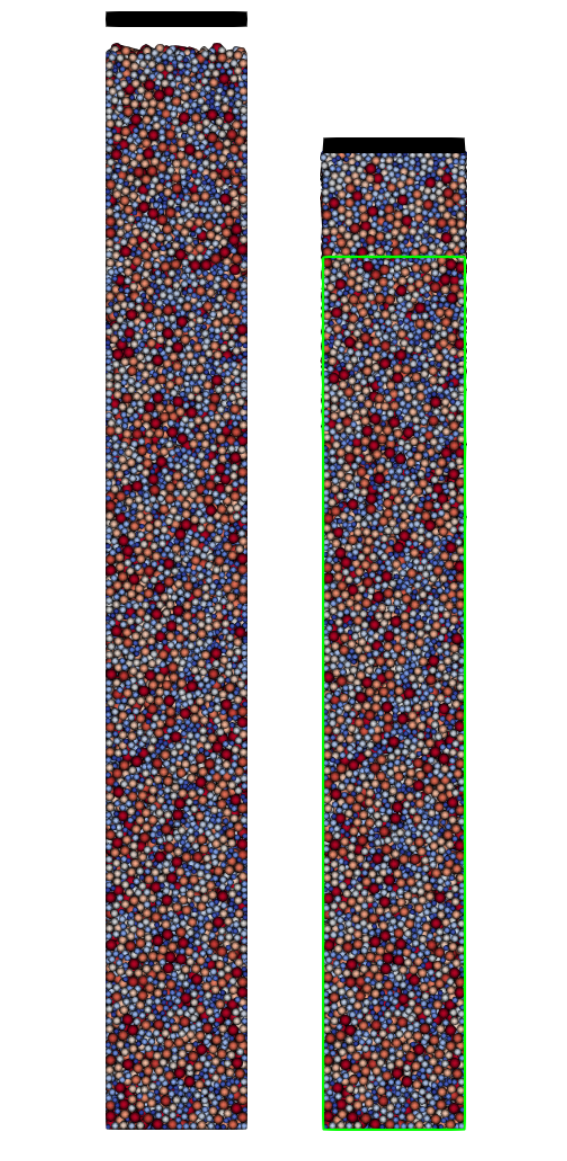}
        \caption{Two steps of the powder compression; the particles inside the green box are extracted and placed inside the cavity.}
        \label{fig:compression}
    \end{subfigure}
    \hfill
    \begin{subfigure}[t]{0.24\textwidth}
        \begin{tikzpicture}
            \begin{scope}
                \node[anchor=south west,inner sep=0] (image) at (0,0) {\includegraphics[width=\textwidth]{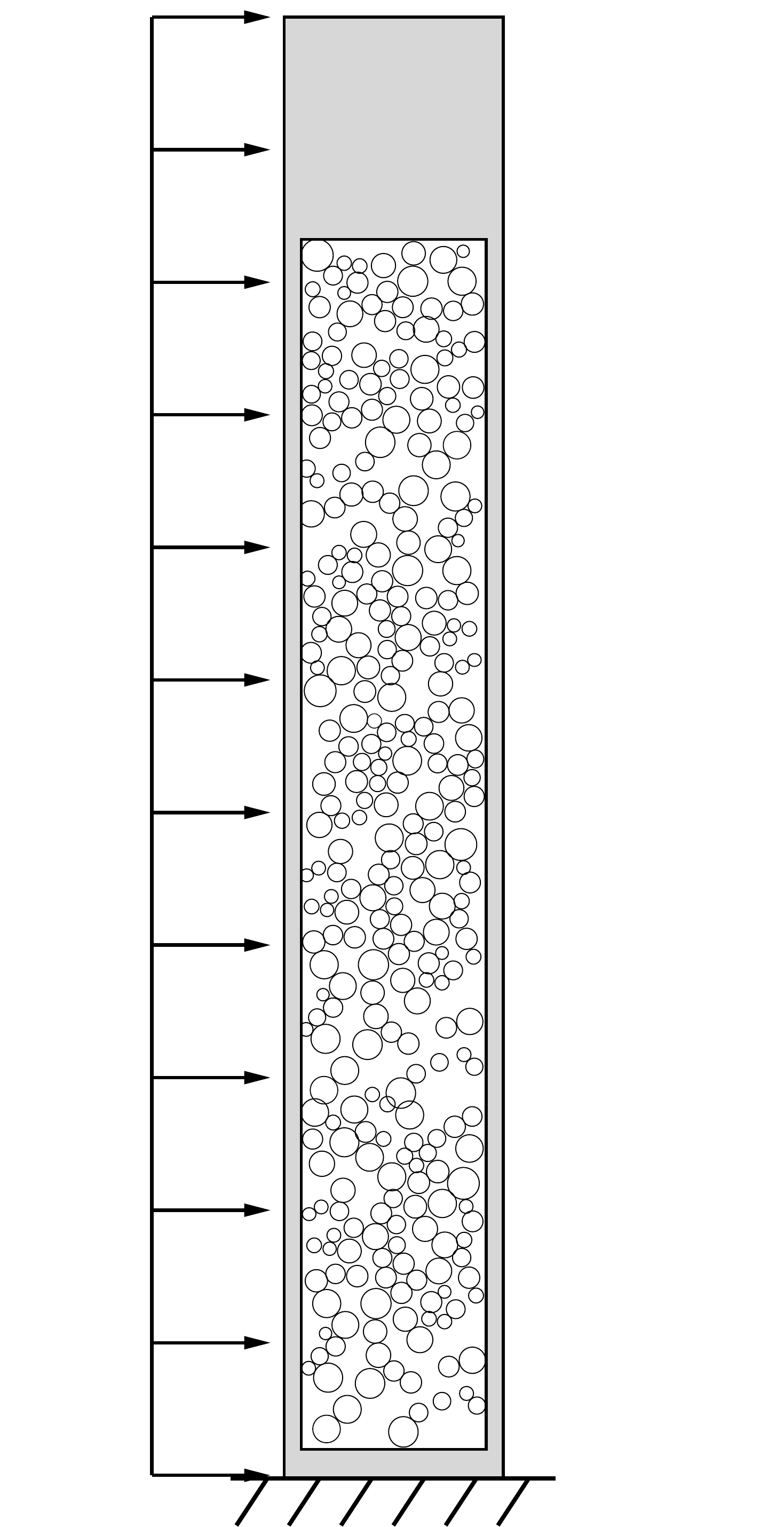}};
                \begin{scope}[x={(image.south east)},y={(image.north west)}]
                    \node at (0.15,0.5) {$q$};
                    \draw[-latex, ultra thick] (0.8,0.8) -- ++(0,-0.2) node[midway,right] {$\mathbf{g}$};
                \end{scope}
            \end{scope}
        \end{tikzpicture}
        \caption{Sketch of the simulation setup.}
        \label{fig:sketch_sim_setup}
    \end{subfigure}
    \hfill
    \begin{subfigure}[t]{0.26\textwidth}
        \includegraphics[width=\textwidth]{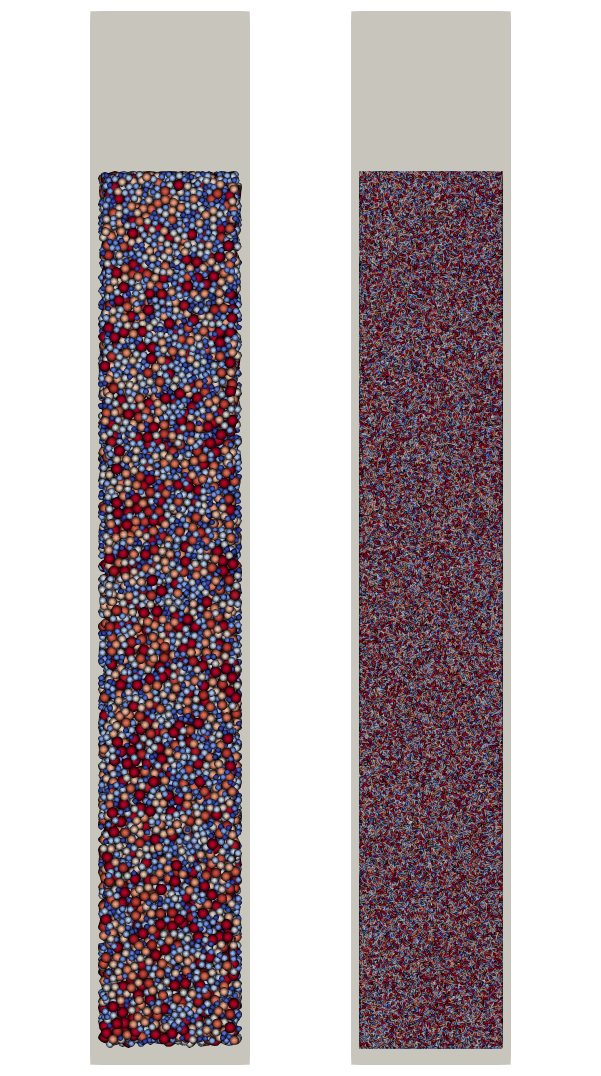}
        \caption{Initial configuration of the beam with cavity for two different powder sizes: $D50 = \SI{1.0}{\milli\meter}$ (left) and $D50 = \SI{0.25}{\milli\meter}$ (right).}
        \label{fig:initial_config}
    \end{subfigure}
    \hfill
    \begin{subfigure}[t]{0.145\textwidth}
        \includegraphics[width=\textwidth]{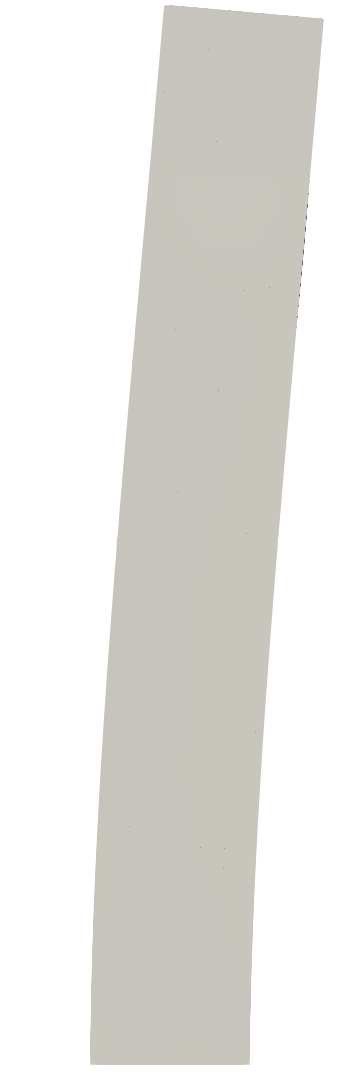}
        \caption{Deflection of the beam (displacement scaled by factor of $100$).}
        \label{fig:excited_beam}
    \end{subfigure}
    \caption{Overview of the simulation setup with powder preparation by powder compression and excitation of a beam with an unfused powder cavity. Outer dimensions $20 \times 20 \times 132 \,\si{\milli\metre\cubed}$ and inner dimensions $18 \times 18 \times 110 \,\si{\milli\metre\cubed}$. The cavity starts~\SI{2}{\milli\meter} above the clamp (Dirichlet fixed bottom surface). The particles are colored according to particle size from smallest (blue) to  largest (red), i.e., $D_{min}~=~\SI{0.71}{\milli\meter}$ to $D_{max}=\SI{1.41}{\milli\meter}$ for D50 = 1.0 mm and $D_{min}=\SI{0.18}{\milli\meter}$ to $D_{max}=\SI{0.35}{\milli\meter}$ for D50 = 0.25 mm.}
    \label{fig:simulation_setup}
\end{figure}

The computational model described above is implemented in the parallel, multi-physics research code BACI~\cite{Baci}. Using this model, cantilever beams with a particle-filled cavity are created to study the effect of enclosed powder particles on the damping of free bending oscillations. The beam has a length of $\SI{132}{\milli\metre}$ with a rectangular cross-section with dimensions $20\times20\,\si{\square\milli\metre}$ (motivated by experiments studied in \cite{Ehlers2021}). The cavity inside the beam has the dimensions $18 \times 18 \times 110 \,\si{\milli\metre\cubed}$, resulting in a wall thickness of~\SI{1}{\milli\meter}. The simulation consists of two steps. In the first step, a random particle configuration is created. In the second step, this particle configuration is used as initial configuration for the parameter study. In order to create a realistic initial configuration for the particles inside the cavity, particles are initially placed on a Cartesian grid and then settle down due to gravity, creating a random powder configuration similar to a configuration resulting from powder spreading. A Dirichlet controlled plate is used to compress the powder, which allows to study the influence of different (pre-defined) packing densities (see Fig.~\ref{fig:compression}). For a desired packing density, the particles are extracted and inserted into the beam cavity for the second simulation step. Figure~\ref{fig:initial_config} shows the initial configuration of the beam with the particle-filled cavity. The clamped end of the beam is realized by a Dirichlet surface condition on the bottom surface of the beam such that the beam is in a vertical position and gravity acts in the downward direction. Initially, the beam is bent by applying a constant area force $q = \SI{0.1}{\newton\per\milli\meter\squared}$ (resultant force $R=\SI{264}{\newton}$) on one side of the beam (see Fig.~\ref{fig:sketch_sim_setup}). The force is linearly increased over~\SI{5}{\milli\second} and then held constant for~\SI{1}{\milli\second} resulting in a maximal transverse tip displacement of approximately~\SI{0.1}{\milli\meter} (see Fig.~\ref{fig:excited_beam}). At $T = \SI{6}{\milli\second}$ the force is removed and the beam begins to oscillate freely.

The particle size follows a log-normal size distribution that was used in our previous work~\cite{Penny2021} with the same type of material (see Table~\ref{tab:parameters_dem}). In the present work a value of $k_N = \SI{5642}{\newton\per\meter}$ has been chosen for the penalty parameter, which limited the maximal particle penetrations to values below~\SI{5}{\percent} of the particle diameter. In Section~\ref{sec:results:contact_stiffness}, the sensitivity of the presented results w.r.t. this choice will be critically analyzed. All parameters of the computational model are summarized in Tables~\ref{tab:parameters_dem} and~\ref{tab:parameters_fem}.
The cavity volume of $18 \times 18  \times 110 \,\si{\milli\meter\cubed}$ filled with particles of original size D50 $\approx \SI{25}{\micro\meter}$ would result in a total of $\num{2.3e9}$ particles. To reduce the computational complexity, the problem size needs to be scaled to make it computationally manageable. Kiracofe et al.~\cite{Kiracofe2021} scaled down the computational domain to only simulate a subset of the particles, but with the original particle size. In this work the original geometry is used and the particle size is upscaled until a manageable amount of particles is reached. Critically, using the original geometry allows a direct comparison with experiments using the same beam geometry. Following this approach, the influence of particle size will be critically studied. For the default setup, a particle size of $D_{50} = \SI{1.0}{\milli\meter}$ is used, which roughly requires \num{32000} particles to fill the cavity and corresponds to a scaling by a factor of~\num{37} relative to the original particle size. A simulated time of $T_{max} = \SI{30}{\milli\second}$ is chosen, which is enough to capture the decrease of the oscillation amplitude. The time step size depends on the particle size~\cite{Meier2019} and is chosen as $\Delta{}t = \SI{2.5}{\micro\second}$ for the particle size $D50 = \SI{1.0}{\milli\meter}$. In all simulations, the structural problem is solved with the same time step size as used for the DEM model.

\section{Results}
\label{sec:results}
Using the computational model described in Section \ref{sec:methods}, the influence of different physical effects on the dissipation is studied, namely the packing density, the penalty parameter of the contact model, the coefficient of friction, the coefficient of restitution, the adhesive forces, the particle size, and finally the orientation of the beam.

For the powder and beam material representative properties of stainless steel 316L are used. The default values of the considered model parameters are summarized in Tables~\ref{tab:parameters_dem} and~\ref{tab:parameters_fem}. The damping behavior is studied based on the decay of the oscillation amplitude. The oscillation is quantified by the displacement of the top center point on the beams outer surface in the direction of the excitation.

\subsection{Packing density}
When manufacturing parts by PBF-LB/M, the powder packing density within closed cavities results from the chosen process parameters. Different process conditions may alter the packing density and ultimately influence the damping capabilities. In the context of this work, the packing density is prescribed in a controlled manner and defined as the ratio of particle volume to cavity volume. The cavity is always completely filled with particles such that there is no clearance between the particle bulk and the cavity walls. The packing density is varied between 50~\% and 60~\% in steps of 1~\%. For these different packing densities, Figure~\ref{fig:packing_fractions_D50_1000um} shows the tip displacement (in x-direction) of the oscillating beam over time. Accordingly, there is no noticeable damping for low packing densities ($\le 52~\%$). For larger packing densities the damping increases. The optimum damping is reached for a packing density of approximately 58~\%, i.e., the amount of damping decreases again when further increasing the packing density. This observation may be explained as follows: at very low packing densities ($\ll 58~\%$) there are too few contacts and the contact normal forces are too small to generate noticeable dissipation. In contrast at very high packing density ($\gg 58~\%$), particles are packed closely together and the associated high contact normal forces hinder slip motion between the particles (i.e., contacting particles predominantly remain in the stick state), leading to reduced dissipation.

\subsection{Particle size}
\label{sec:results:particle_size}
\begin{figure}[htb]
    \centering
    \begin{subfigure}{0.45\textwidth}
        \includegraphics[width=\textwidth]{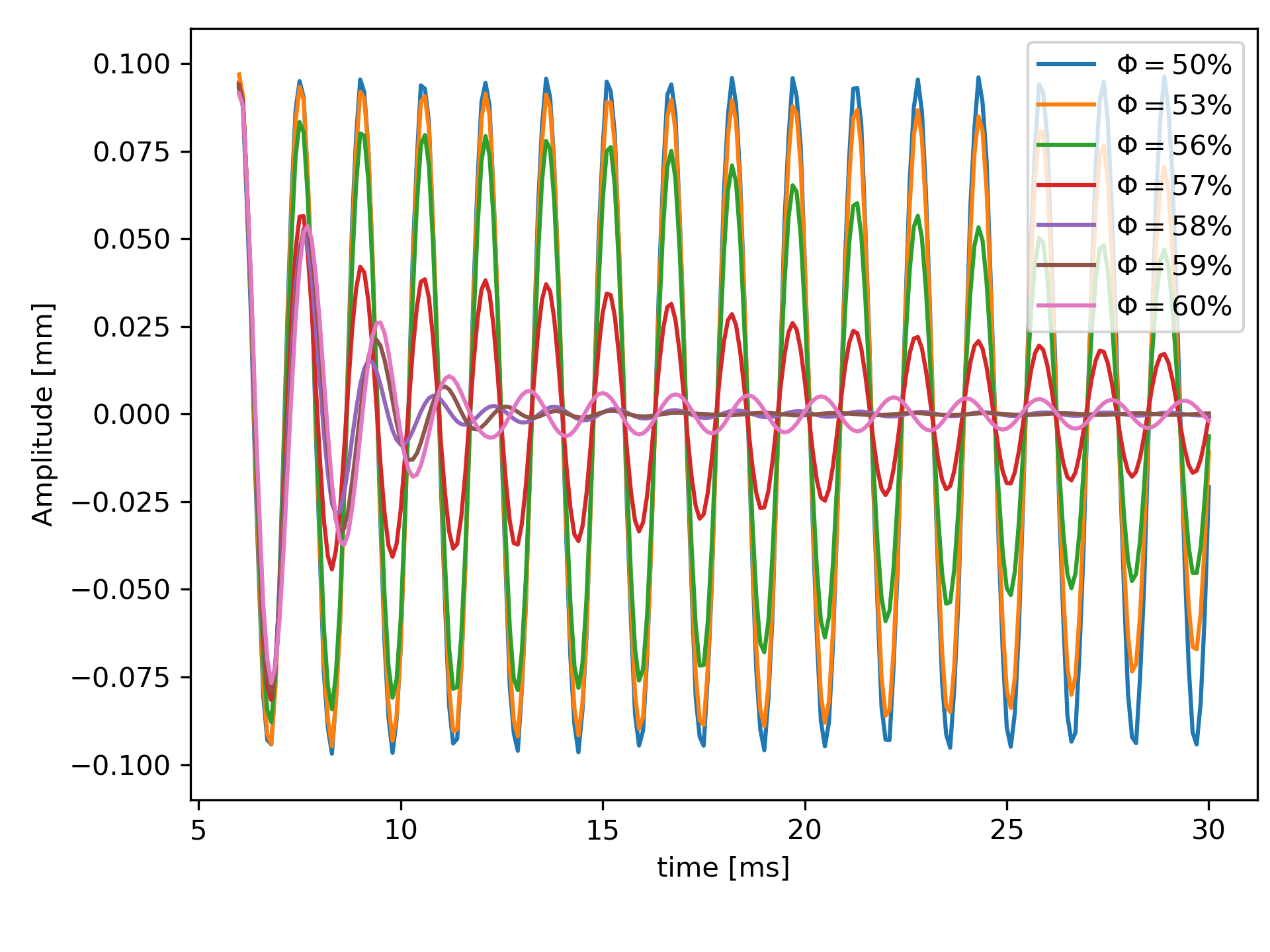}
        \caption{D50 = 1.0 mm.}
        \label{fig:packing_fractions_D50_1000um}
    \end{subfigure}
    \begin{subfigure}{0.45\textwidth}
        \includegraphics[width=\textwidth]{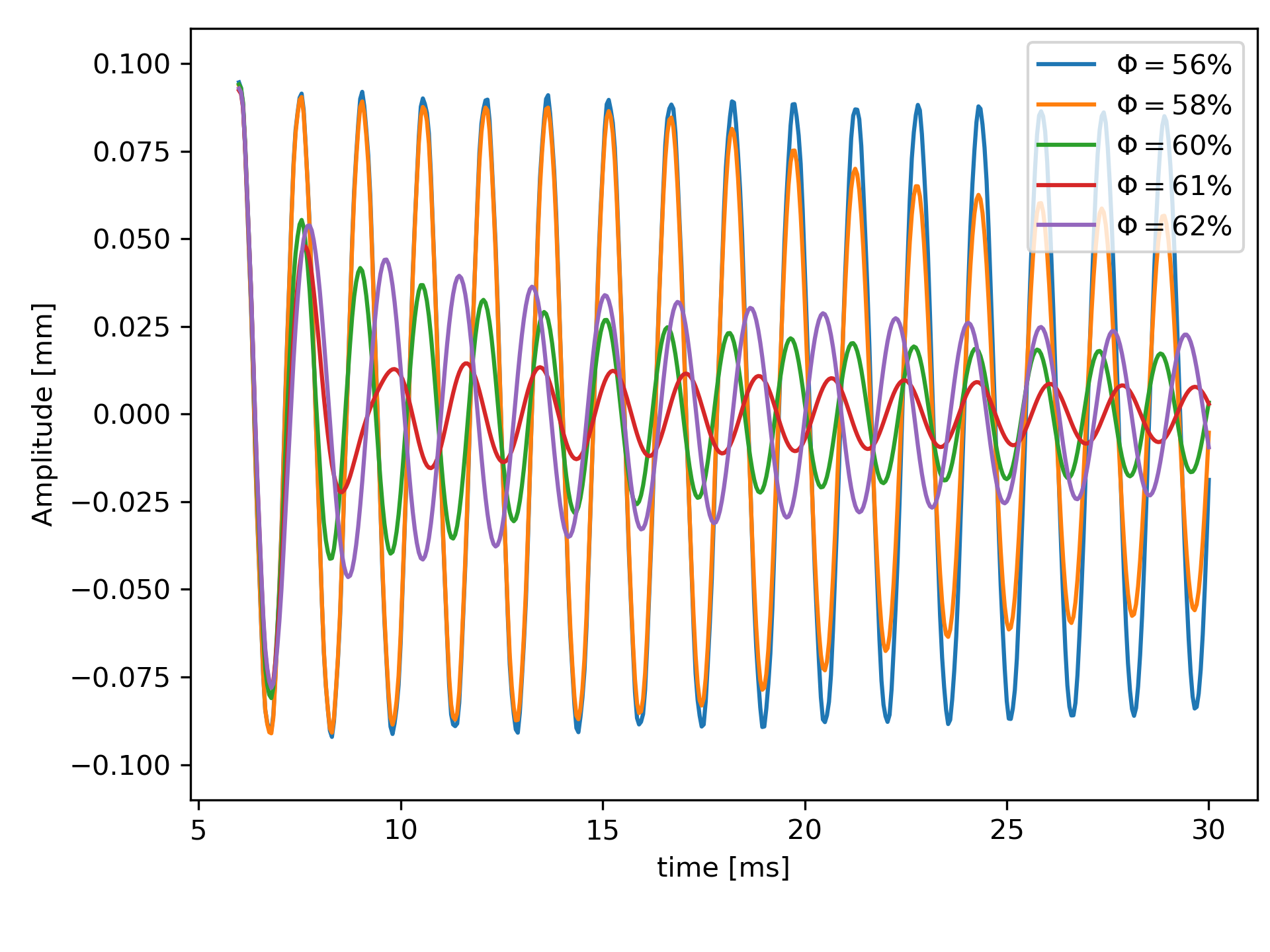}
        \caption{D50 = 0.5 mm.}
        \label{fig:packing_fractions_D50_500um}
    \end{subfigure}
    \begin{subfigure}{0.45\textwidth}
        \includegraphics[width=\textwidth]{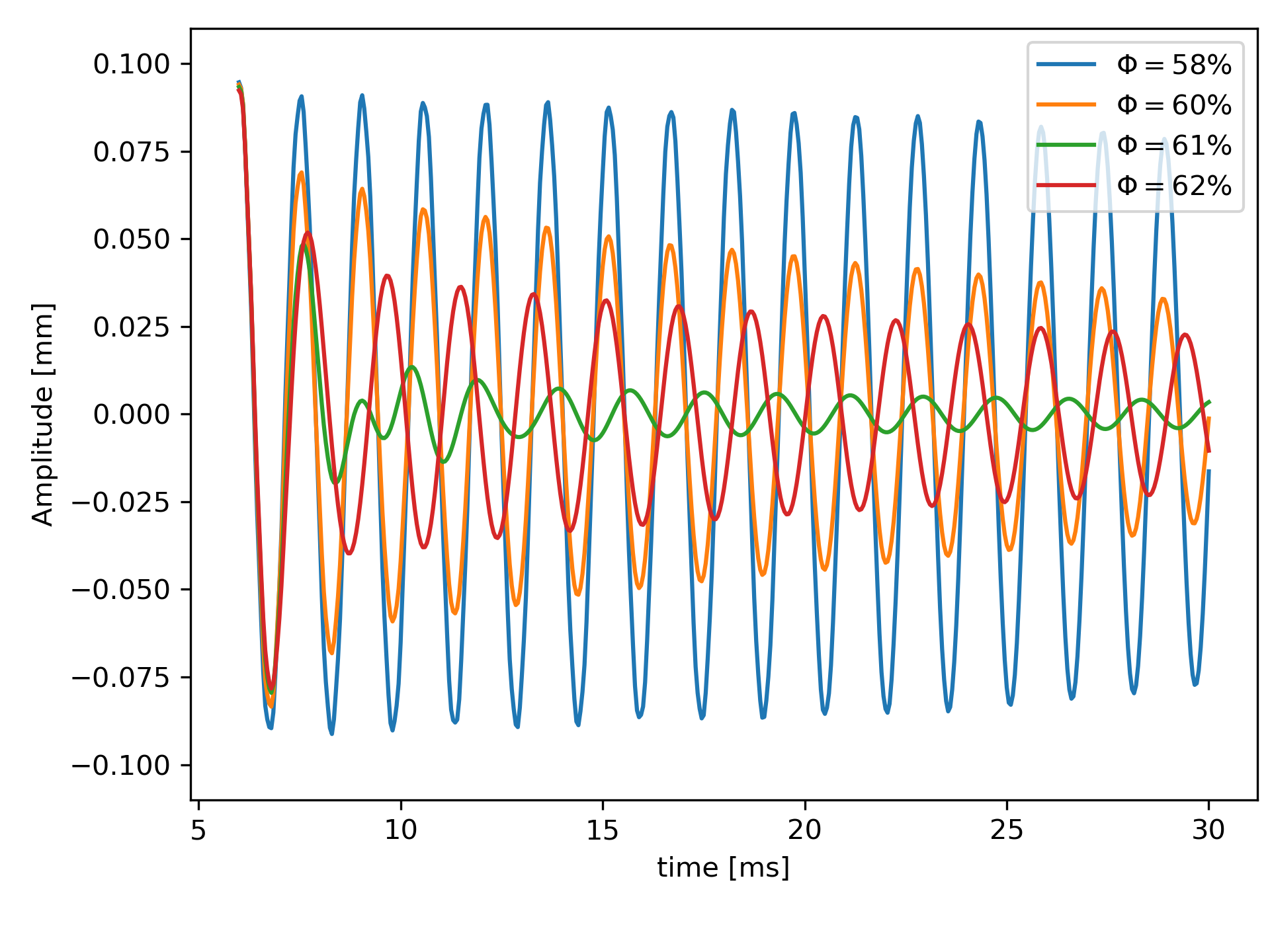}
        \caption{D50 = 0.25 mm.}
        \label{fig:packing_fractions_D50_250um}
    \end{subfigure}
    \caption{Influence of different packing densities $\Phi$ on the amplitude.}
    \label{fig:packing_fractions}
\end{figure}

To study the effect of the particle size, the default particle size of~\SI{1}{\milli\meter} is scaled by a factor of 0.5 and 0.25, respectively. Additionally, different packing densities for each particle size are simulated.  When the particle size is reduced by a factor of \num{2} to $D_{50}=\SI{0.5}{\milli\meter}$ with packing densities ranging from $\Phi=\SI{56}{\percent}$ to~\SI{62}{\percent} ($\sim{}250000$ particles), an optimum packing density can be identified at $\Phi=\SI{61}{\percent}$ (Figure~\ref{fig:packing_fractions_D50_500um}). However, the damping is slightly less than for the default case. Another reduction in particle size by a factor of \num{2}, i.e., $D_{50}=\SI{0.25}{\milli\meter}$ ($\sim{}2.0$ million particles), yields similar results (Figure~\ref{fig:packing_fractions_D50_250um}). Again, the best damping is obtained for $\Phi=\SI{61}{\percent}$. 

To get a sense for the damping capabilities, the damping ratio for the particle sizes $D50 = 0.25$, $0.5$, and $1.0~\si{\milli\meter}$ are summarized in Table~\ref{tab:damping_ratio} for the respective optimal packing densities. As damping is highly non-linear for particle dampers, the damping ratio of the first two periods is given, starting from the first local minimum, and the damping ratio for the fifth period, computed from the first to the sixth local minimum. The damping ratio $\zeta = 1 \big/ \sqrt{1+ \left(2\pi / \delta \right)^2}$ is calculated from the logarithmic decrement $\delta = \frac{1}{n} \ln\left(x(t) / x(t+nT)\right)$, with the oscillation period T and the integer $n$ (where $n=1$ for $\zeta_1$, $\zeta_2$ and $n=5$ for $\bar{\zeta}_5$). The damping ratio at the respective optimal packing densities is quite similar. The damping ratios at the optimal packing densities of the different particle sizes may even be closer together when more packing densities are studied near the current optimum. Generally, the particle size has only a small influence on the dissipation behavior compared to the large influence of the packing density.

\begin{table}
    \centering
    \caption{Damping ratio for different particles sizes and their respective optimal packing density. $\zeta_i$ is the damping ratio for the $i$-th period where the first period is chosen from the first to the second local minimum (see Fig.~\ref{fig:packing_fractions}). $\bar{\zeta}_5$ is the damping ratio calculated over five periods, i.e., from the first to the sixth local minimum.}
    \label{tab:damping_ratio}
    \begin{tabular}{lllll}
        \toprule
        Particle size D50 [mm] & Packing density $\Phi$ [\si{\percent}] & $\zeta_1$ & $\zeta_2$ & $\bar{\zeta}_5$ \\
        \midrule
        0.25 & 61 & 0.217 & 0.166 & 0.075 \\
        0.5  & 61 & 0.196 & 0.059 & 0.060 \\
        1.0  & 58 & 0.160 & 0.186 & 0.121 \\
        \bottomrule        
    \end{tabular}
\end{table}

\subsection{Contact stiffness}
\label{sec:results:contact_stiffness}
\begin{figure}[htb]
    \centering
    \begin{subfigure}{0.45\textwidth}
        \includegraphics[width=\textwidth]{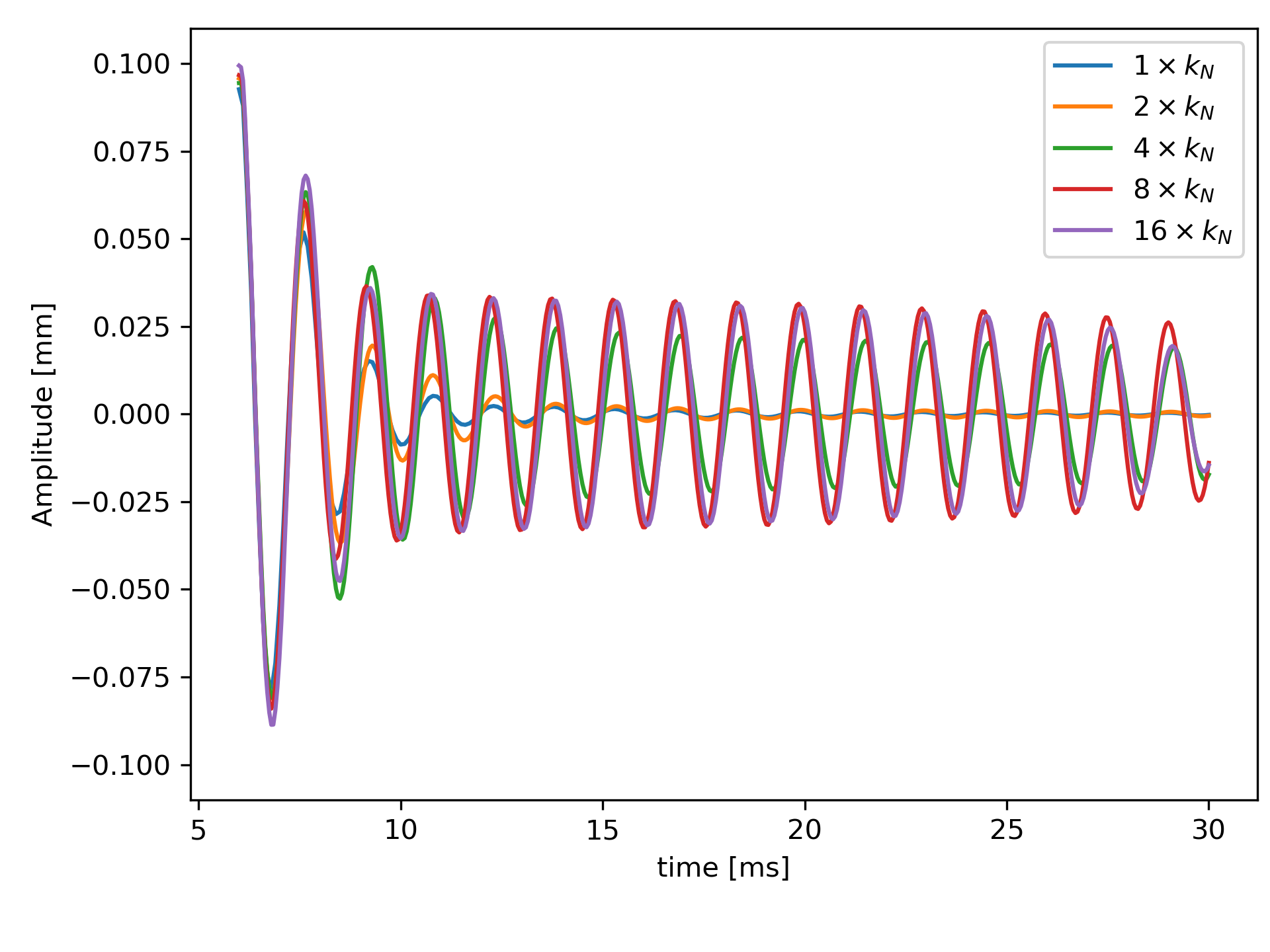}
        \caption{Different contact stiffnesses for $\Phi=58~\%$.}
        \label{fig:stiffness_variation}
    \end{subfigure}
    \begin{subfigure}{0.45\textwidth}
        \includegraphics[width=\textwidth]{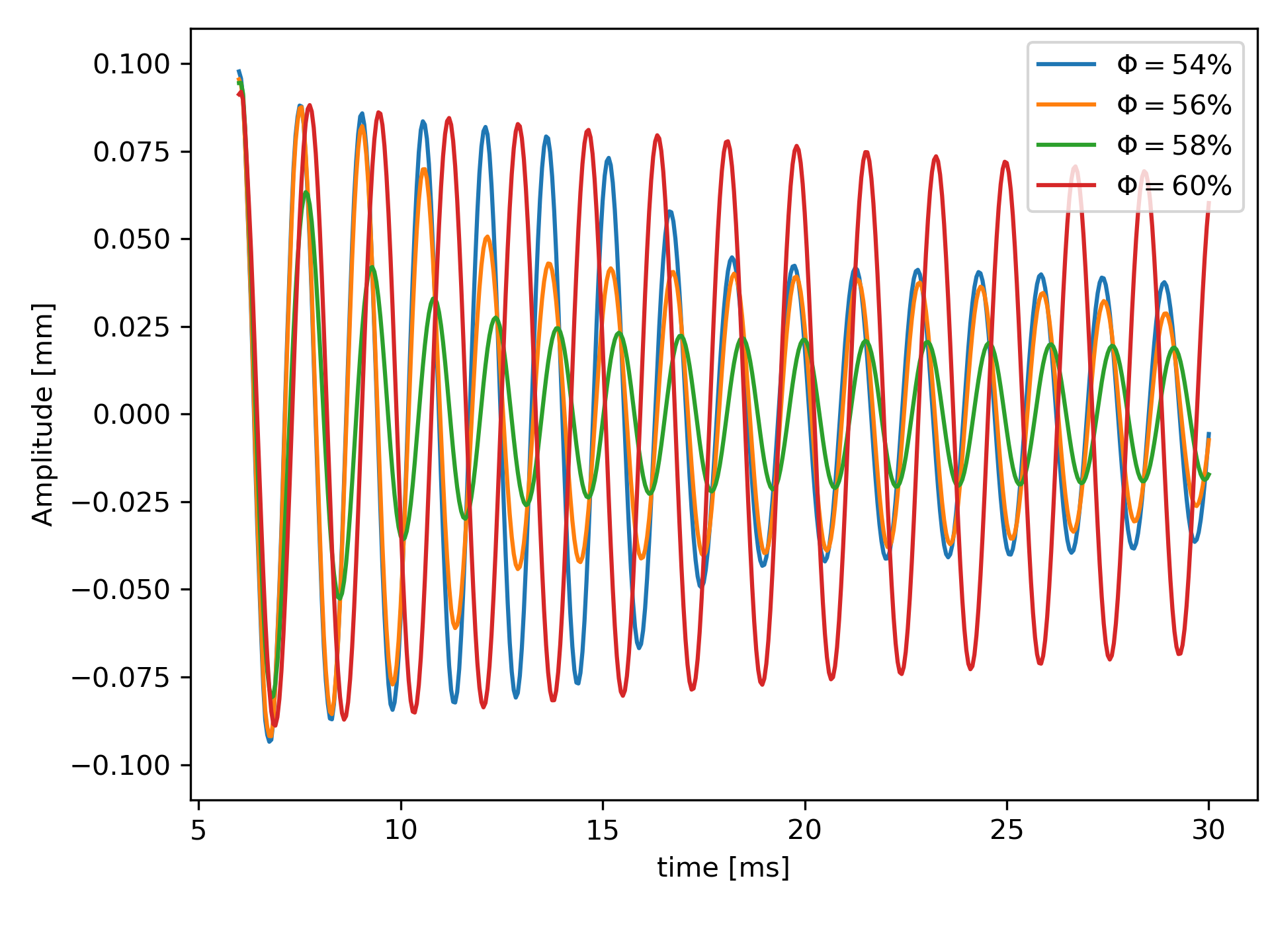}
        \caption{Different packing densities for contact stiffness $4 \times k_N$.}
        \label{fig:stiffness_packing_fraction}
    \end{subfigure}
    \caption{Influence of contact stiffness $k_N$ on the amplitude.}
    \label{fig:contact_stiffness}
\end{figure}

As typical for DEM simulations, the contact stiffness is intentionally chosen lower than the actual Young's modulus of the material to allow for significantly larger time step sizes. To check the validity of the stiffness reduction, additional damping simulations are performed with increased values of the penalty parameter. Figure~\ref{fig:stiffness_variation} shows the oscillations of the beam with packing density $\Phi = \SI{58}{\percent}$ for the default contact stiffness $k_N = \SI{5642}{\newton\per\meter}$ and increased values by a factor of 2, 4, 8, and 16, respectively. Accordingly, increasing the contact stiffness leads to a reduced dissipation. However, at a penalty value of approximately $4 \times k_N$ a saturation is observed, i.e., further increasing the penalty parameter does not alter the damping behavior anymore. To check the detailed influence of the increased contact stiffness (by a factor 4), the packing density is varied from~\SIrange{50}{60}{\percent} for this case. Figure~\ref{fig:stiffness_packing_fraction} shows the displacement curve for selected packing densities. Also for the increased contact stiffness, there is a strong dependence of the dissipation characteristics on the packing density and an optimal dissipation is achieved at $\Phi=58~\%$. Thus, it is concluded that the same fundamental trends and correlations already observed for the original penalty parameter $k_N$ are also visible for higher penalty values, which are closer to the actual stiffness characteristics of metallic powders. Consequently, the (computationally cheaper) default contact stiffness $k_N$ will be applied in the remainder of this work.

\subsection{Coefficient of friction}
With friction being one of the two main dissipation mechanisms for particle dampers, the coefficient of friction is varied from $\mu = 0$ to $\mu = 0.7$ (in steps of 0.1), while the coefficient of restitution is $e=0.4$. According to Fig.~\ref{fig:friction}, coefficients of friction $\mu \leq 0.2$ show only moderate damping. In particular, this behavior can also be observed when no friction is considered at all. This suggests that a certain portion of the dissipation can be attributed to inelastic impacts. The damping increases for higher values of the coefficient of friction (as consequence of higher sliding friction forces) and reaches an optimum for $\mu = 0.4$. Further increasing the coefficient of friction slightly reduces the damping again. A greater coefficient of friction results in greater transferable tangential forces. So, similar to the observations already made for very high packing fractions (i.e., high contact normal forces), for a greater coefficient of friction a significant portion of particles switches from slip to stick friction, reducing the overall dissipation.

\subsection{Coefficient of restitution}
\begin{figure}[htb]
    \centering
    \begin{subfigure}{0.45\textwidth}
        \includegraphics[width=\textwidth]{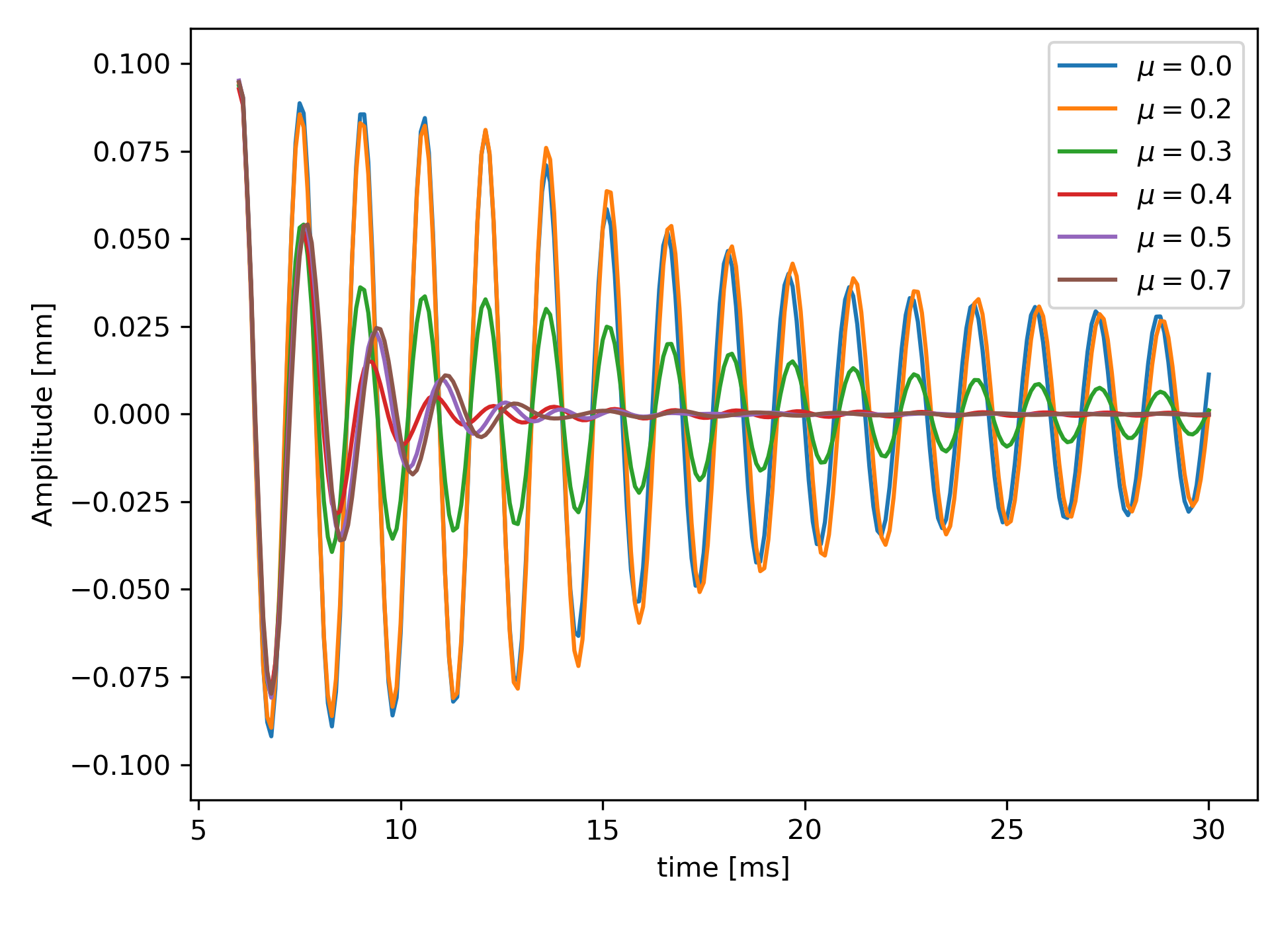}
        \caption{Coefficient of friction $\mu$ ($e=0.4$).}
        \label{fig:friction}
    \end{subfigure}
    \begin{subfigure}{0.45\textwidth}
        \includegraphics[width=\textwidth]{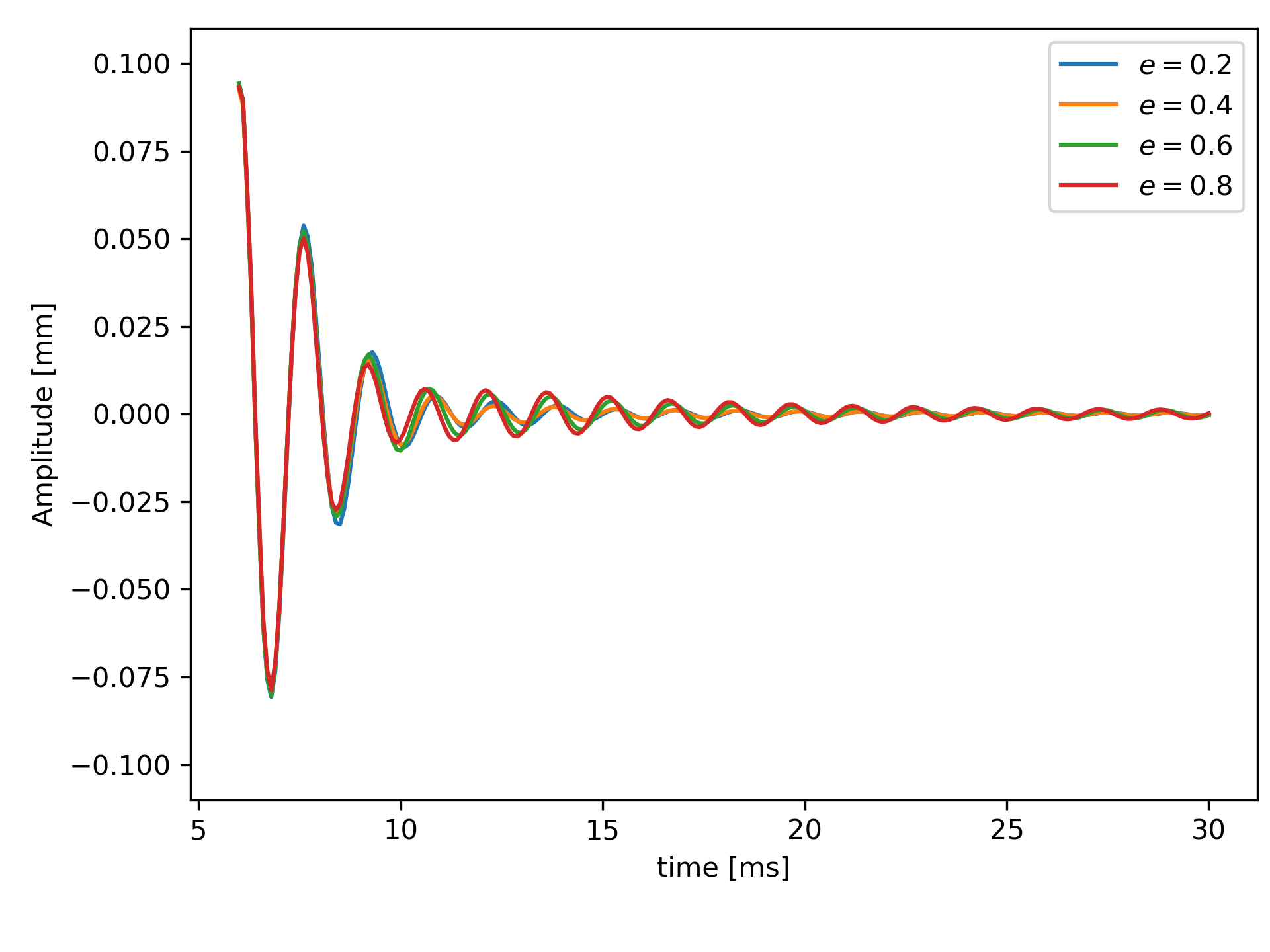}
        \caption{Coefficient of restitution $e$ ($\mu = 0.4$).}
        \label{fig:restitution}
    \end{subfigure}
    \caption{Influence of coefficient of friction $\mu$ and coefficient of restitution $e$ on the amplitude.}
    \label{fig:friction_and_restitution}
\end{figure}

Varying the coefficient of restitution in the range from $0.2$ to $0.8$ (while keeping the coefficient of friction at the default value of $\mu=0.4$) shows an almost identical dissipation behavior (see Fig.~\ref{fig:restitution}). Thus, it is considered that the influence of the coefficient of restitution is negligible for the studied range of packing densities and friction is the dominating form of dissipation. Though, impacts also contribute to energy dissipation up to a certain degree, as shown in Fig.~\ref{fig:friction}, where frictionless contact ($\mu=0$) still leads to some reduction of the oscillation amplitude.

\subsection{Adhesion}
\begin{figure}[htb]
    \centering
    \begin{subfigure}{0.45\textwidth}
        \includegraphics[width=\textwidth]{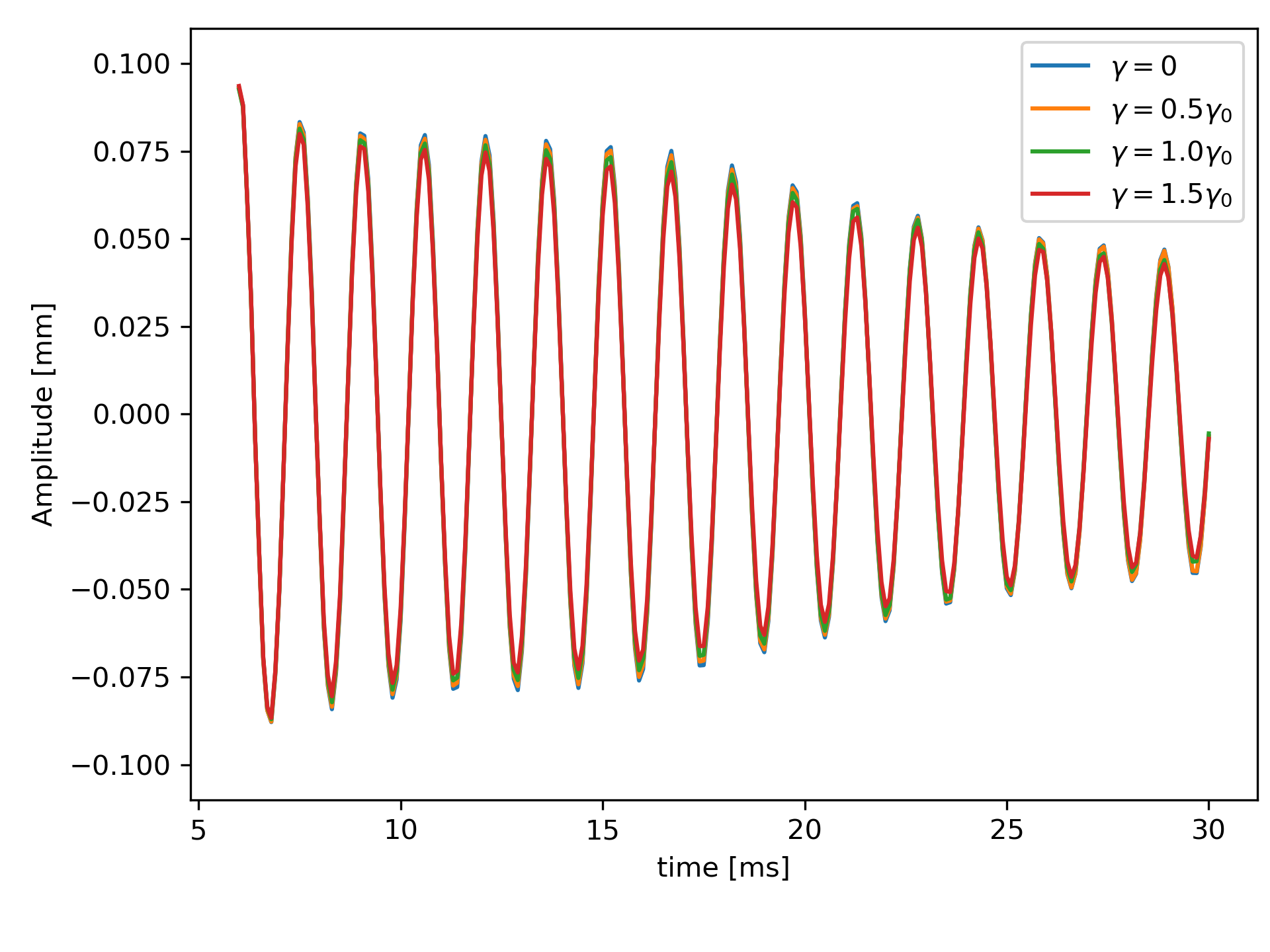}
        \caption{$\Phi = 56~\%$.}
    \end{subfigure}
    \begin{subfigure}{0.45\textwidth}
        \includegraphics[width=\textwidth]{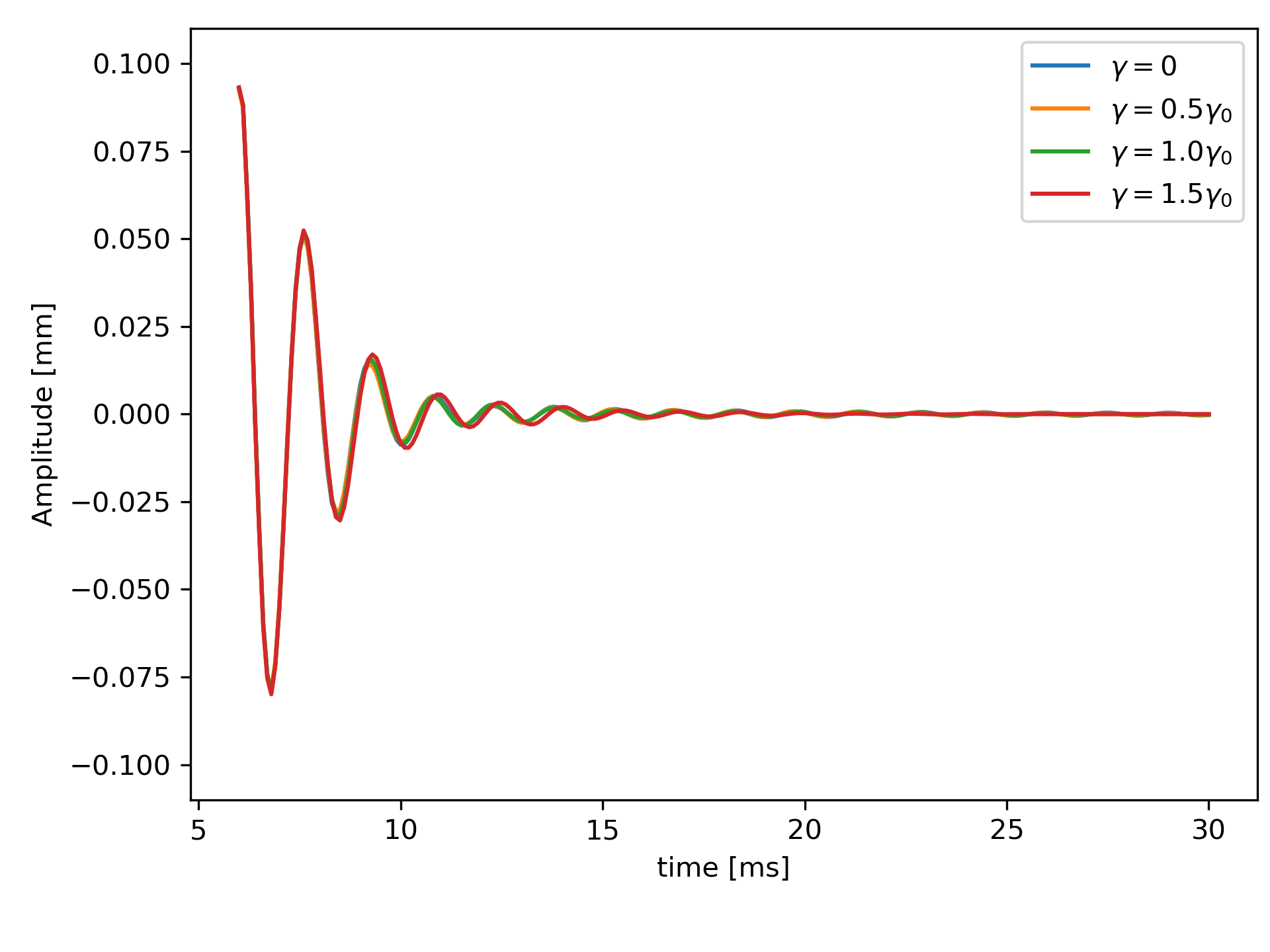}
        \caption{$\Phi = 58~\%$.}
    \end{subfigure}
    \caption{Influence of adhesion (with the scaled original surface energy $\gamma_0$).}
    \label{fig:adhesion}
\end{figure}

For small powder particles, as used in additive manufacturing, adhesive forces are known to affect the flow behavior of bulk powder significantly. For example, reducing the particle size by a factor of 2, reduces the gravitational force by a factor of 8 while the adhesive forces reduce only by a factor of 4. Therefore, adhesive forces dominate over gravitational forces for small particle sizes and have a major influence on the powder flowability, e.g., as apparent in powder spreading~\cite{Meier2019, Meier2019a}. Now, the influence of adhesive forces, which were not considered in the previous results, are studied. To account for the up-scaled particle size, the cohesive surface energy is upscaled by the square of the scaling factor to keep the dimensionless powder cohesiveness, i.e., the ratio of gravity to adhesion forces unchanged~\cite{Meier2019}. Accordingly, the surface energy of the original powder $\gamma_{ref} = \SI{0.06}{\milli\joule\per\meter\squared}$ was scaled to $\gamma_{scaled} = \SI{85}{\milli\joule\per\meter\squared} = \gamma_0$ to account for the upscaled powder particle size. To investigate the general influence of cohesion, the surface energy values $0.5\cdot\gamma_{scaled}$ and $2\cdot\gamma_{scaled}$ are studied in addition. Figure~\ref{fig:adhesion} shows that adhesive forces and different magnitudes thereof do not significantly influence the damping. Thus, it is assumed that the normal forces are mainly a result of the pre-compression step resulting in rather high packing density such that the adhesive forces only have a minor contribution.

\subsection{Beam orientation}
In the default case the beam is oriented vertically with the clamped end at the bottom. In additional simulations the beam is oriented upside down, i.e., vertically with the clamped end at the top, and horizontally with horizontal excitation. This is easily achieved in the simulations by changing the direction of gravity. The different setups are studied for two packing densities.

\begin{figure}[htb]
    \centering
    \begin{subfigure}{0.45\textwidth}
        \includegraphics[width=\textwidth]{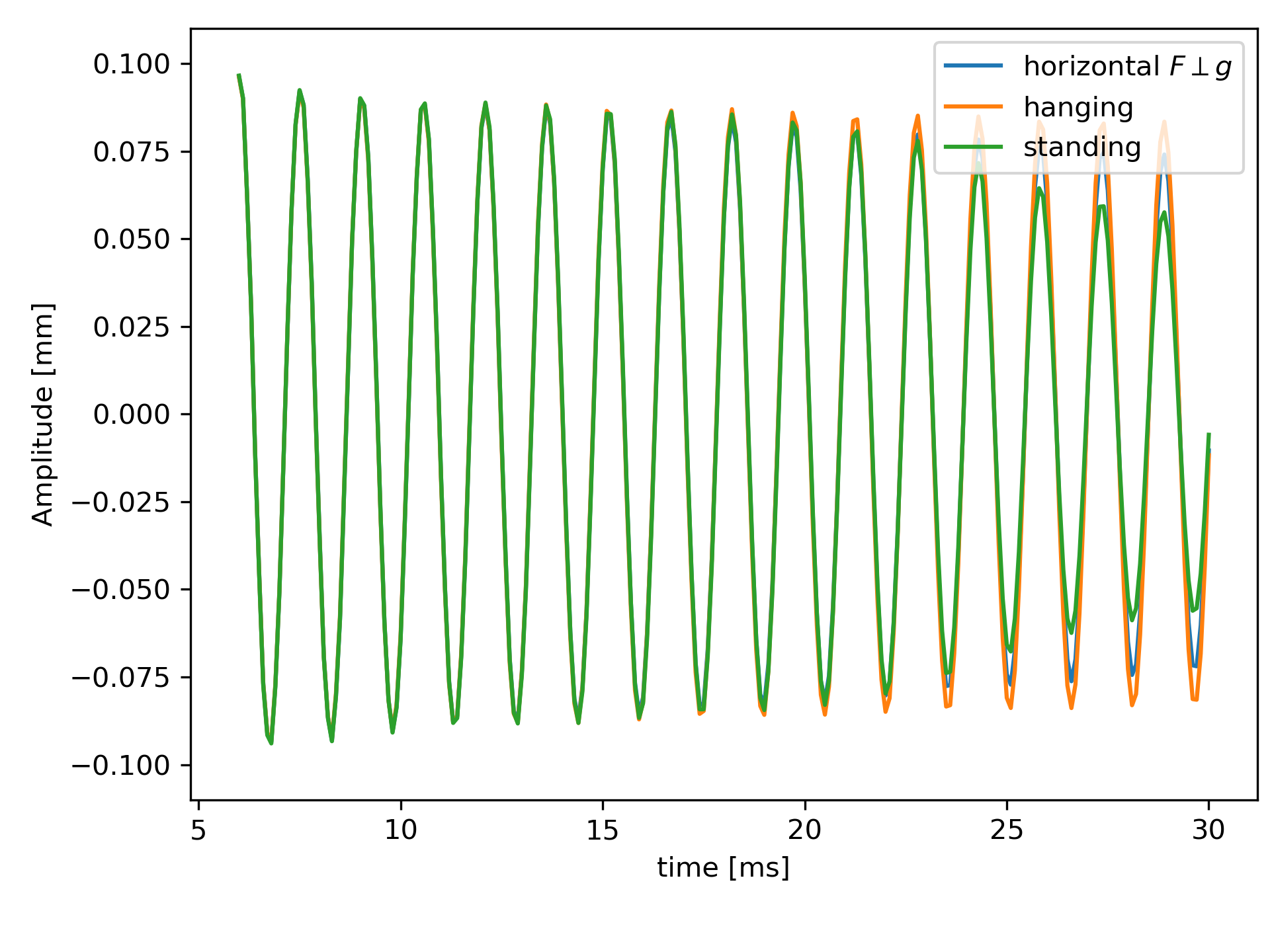}
        \caption{$\Phi = 54~\%$.}
    \end{subfigure}
    \begin{subfigure}{0.45\textwidth}
        \includegraphics[width=\textwidth]{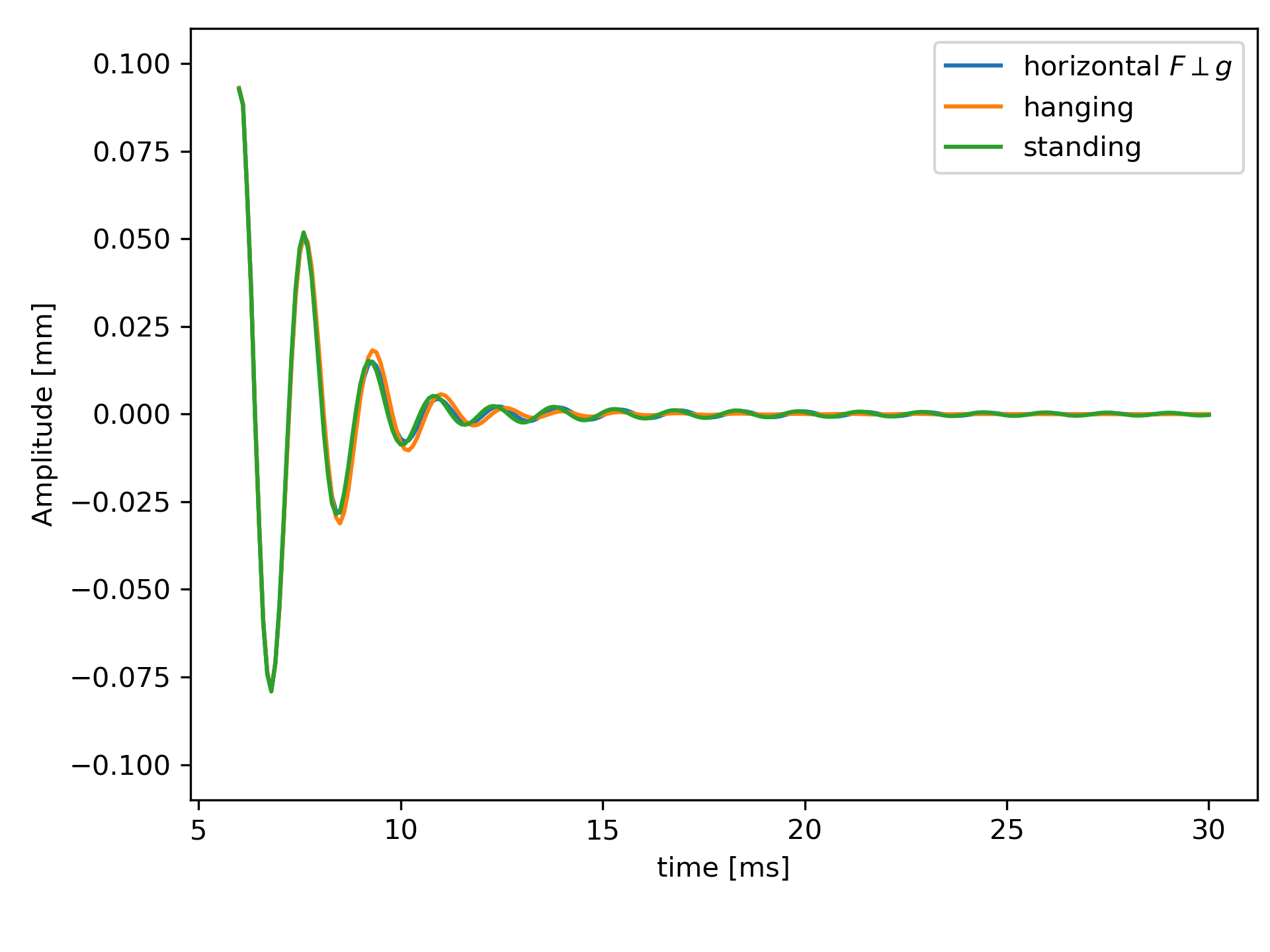}
        \caption{$\Phi = 58~\%$.}
    \end{subfigure}
    \caption{Influence of the beam orientation.}
    \label{fig:gravity}
\end{figure}

For the default packing density of $\Phi=58~\%$ a different orientation does not influence the damping (Fig.~\ref{fig:gravity}b). Due to the high packing density there is not enough free motion possible for the particles to rearrange when turning the beam. In contrast to that, the packing density $\Phi=54~\%$ allows for a more significant particle reconfiguration (Fig.~\ref{fig:gravity}a) such that the beam orientation has a visible, yet small, influence on the damping.

\section{Analytical study on the influence of the particle size}
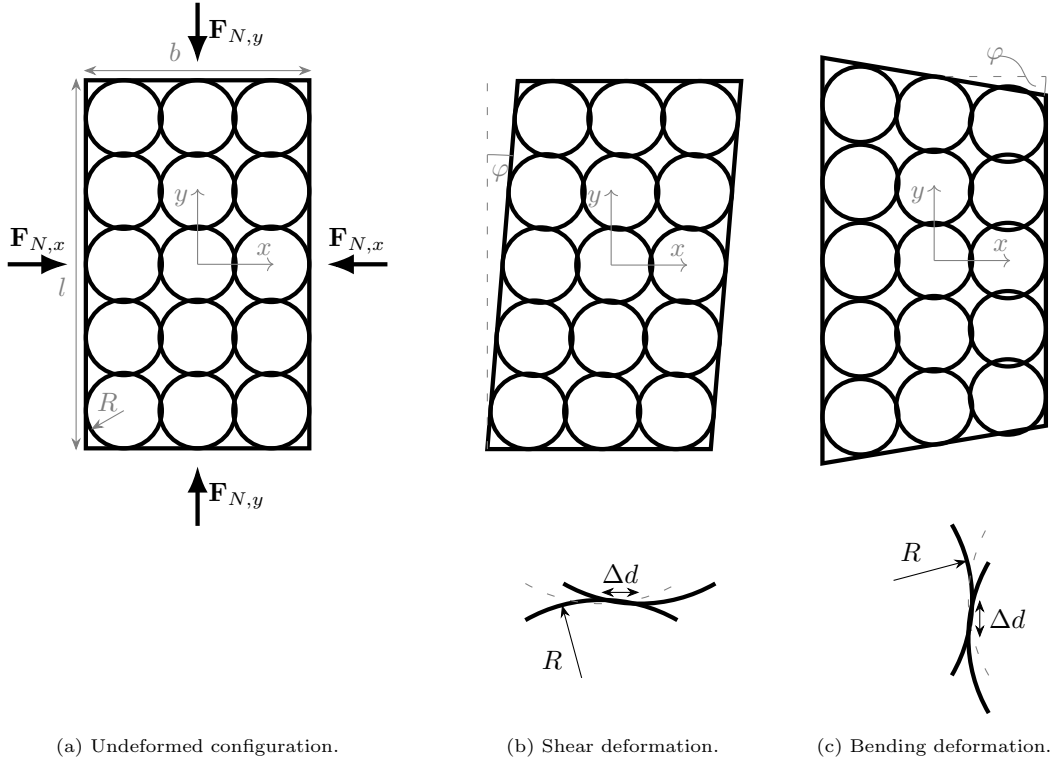
\begin{figure}
    \centering
    \begin{subfigure}[t]{0.4\textwidth}
        \centering
            \begin{tikzpicture}[scale=1, every node/.style={scale=1}]
        \def\gap{0.03}
        \def\radius{0.5}
        \def\nx{3}
        \def\ny{5}
        \pgfmathsetmacro\l{2*\radius*\ny-\gap*(\ny-1)} 
        \pgfmathsetmacro\b{2*\radius*\nx-\gap*(\nx-1)}
        \begin{scope}
            \node[draw, ultra thick, rectangle, minimum width={\b cm}, minimum height={\l cm}](default_rect) at ({\b*0.5},{\l*0.5}) {};
            \foreach \discardi [count=\i from 0] in {1,...,\nx}
                \foreach \discardj [count=\j from 0] in {1,...,\ny}
                {
                    \draw[ultra thick] ($ (\radius,\radius)+{2*\radius-\gap}*(\i,\j)$) circle (\radius);
                }
            \draw[->, thin, gray] ({\b*0.5},{\l*0.5}) -- ++(1,0) node[very near end, above] {$x$};
            \draw[->, thin, gray] ({\b*0.5},{\l*0.5}) -- ++(0,1) node[very near end, left] {$y$};
            \draw[-Stealth, thin, gray] (\radius,\radius) -- node[midway,anchor=south]{$R$} ++(210:\radius cm);
            \draw[Stealth-Stealth, thin, gray] ([yshift=0.1cm]default_rect.north west) -- ([yshift=0.1cm]default_rect.north east) node[midway, above, xshift=-0.3cm]{$b$};
            \draw[Stealth-Stealth, thin, gray] ([xshift=-0.1cm]default_rect.north west) -- ([xshift=-0.1cm]default_rect.south west) node[midway, left, yshift=-0.3cm]{$l$};
            \draw[{Latex}-, ultra thick] ([xshift=-0.2cm]default_rect.west) -- ++(-0.8,0) node[midway,anchor=south]{$\mathbf{F}_{N,x}$};
            \draw[{Latex}-, ultra thick] ([xshift=0.2cm]default_rect.east) -- ++(0.8,0) node[midway,anchor=south]{$\mathbf{F}_{N,x}$};
            \draw[{Latex}-, ultra thick] ([yshift=0.2cm]default_rect.north) -- ++(0,0.8) node[midway,anchor=west]{$\mathbf{F}_{N,y}$};
            \draw[{Latex}-, ultra thick] ([yshift=-0.2cm]default_rect.south) -- ++(0,-0.8) node[midway,anchor=west]{$\mathbf{F}_{N,y}$};
            \node at (0,-3.5) {};
        \end{scope}
    \end{tikzpicture}%
        \caption{Undeformed configuration.}
        \label{fig:analytical_model_undeformed}     
    \end{subfigure}
    \begin{subfigure}[t]{0.25\textwidth}
        \centering
            \begin{tikzpicture}[scale=1, every node/.style={scale=1}]
        \def\gap{0.03}
        \def\radius{0.5}
        \def\nx{3}
        \def\ny{5}
        \pgfmathsetmacro\l{2*\radius*\ny-\gap*(\ny-1)} 
        \pgfmathsetmacro\b{2*\radius*\nx-\gap*(\nx-1)}
        \node[draw, draw opacity=0, rectangle, minimum width=3.5cm, minimum height=9.2cm] at (1.65,1.0) {};
        \begin{scope}
            \def\shear{0.4}
            \coordinate (bottomleft) at (0,0);
            \coordinate (topleft) at ($ (0+\shear,\l) $);
            \coordinate (topright) at ($ (\b+\shear,\l) $);
            \coordinate (bottomright) at ($ (\b,0) $);
            \draw[ultra thick] (bottomleft) -- (topleft) -- (topright) -- (bottomright) -- cycle;
            \def\dx{0.075}
            \pgfmathsetmacro\dx{(\shear/\l*(2*\radius-\gap)}
            \foreach \discardi [count=\i from 0] in {1,...,\nx}
                \foreach \discardj [count=\j from 0] in {1,...,\ny}
                {
                    \draw[ultra thick] ($ (\radius+\shear/\l*\radius,\radius)+{2*\radius-\gap}*(\i,\j) +\dx*(\j,0)$) circle (\radius);
                }
            \draw[->, thin, gray] ({\b*0.5+2*\dx},{\l*0.5}) -- ++(1,0) node[very near end, above] {$x$};
            \draw[->, thin, gray] ({\b*0.5+2*\dx},{\l*0.5}) -- ++(0,1) node[very near end, left] {$y$};
            \draw[loosely dashed, thin, gray] (bottomleft) -- ++(90:\l);
            \draw[thin, gray] (0,0.8*\l) arc[start angle = 90, end angle=85, radius=0.8*\l] node[midway, below] {$\varphi$};
        \end{scope}
        \begin{scope}[shift={(1.5,-4)}]
            \draw[ultra thick] (120:2cm) arc (120:60:2cm) node[near start](temp_for_radius) {};
            \draw[loosely dashed, thin, gray] ($ (240:2cm) + (0,3.95) $) arc (240:300:2cm) node [midway](contact_old) {} ;
            \draw[ultra thick] ($ (240:2cm) + (0.5,3.95) $) arc (240:300:2cm) node[midway](contact_new) {};
            \draw[Stealth-] (temp_for_radius.center) -- ($ (temp_for_radius.center)!.5!(0,0) $) node[midway, below left] {$R$};
            \draw[Stealth-Stealth] ([yshift=0.15cm]contact_old.center) -- ([yshift=0.15cm]contact_new.center) node[midway, above] {$\Delta{}d$};
        \end{scope}
    \end{tikzpicture}%
        \caption{Shear deformation.}
        \label{fig:analytical_model_shear}     
    \end{subfigure}
    \begin{subfigure}[t]{0.25\textwidth}
        \centering
            \begin{tikzpicture}[scale=1, every node/.style={scale=1}]
        \def\gap{0.03}
        \def\radius{0.5}
        \def\nx{3}
        \def\ny{5}
        \pgfmathsetmacro\l{2*\radius*\ny-\gap*(\ny-1)} 
        \pgfmathsetmacro\b{2*\radius*\nx-\gap*(\nx-1)}
        \node[draw, draw opacity=0, rectangle, minimum width=3.2cm, minimum height=9.2cm] at (0,-1.5) {};
        \begin{scope}
            \def\ymax{0.25}
            \coordinate (bottomleft) at ($ (-0.5*\b, -0.5*\l-\ymax) $);
            \coordinate (topleft) at ($ (-0.5*\b, +0.5*\l+\ymax) $);
            \coordinate (topright) at ($ (0.5*\b, 0.5*\l-\ymax) $);
            \coordinate (bottomright) at ($ (0.5*\b, -0.5*\l+\ymax) $);
            \draw[ultra thick] (bottomleft) -- (topleft) -- node[midway](boxtop) {} (topright) node (topright_deformed) {} -- (bottomright) -- cycle;
            \foreach \i in {-1.0, 0.0, 1.0}
                \foreach \j in {-2.0,-1.0,0.0,1.0,2.0}
                {
                    \pgfmathsetmacro\x{0+(2*\radius-\gap)*\i}
                    \pgfmathsetmacro\y{0+(2*\radius-\gap)*\j}
                    \draw[ultra thick] (\x, \y -\ymax*4/\b/\l*\x*\y) circle (\radius);
                }
            \draw[->, thin, gray] (0,0) -- ++(1,0) node[very near end, above] {$x$};
            \draw[->, thin, gray] (0,0) -- ++(0,1) node[very near end, left] {$y$};
            \draw[loosely dashed, thin, gray](boxtop) -- ++(0:0.5*\b) node (topright_undeformed) {};
            \draw[thin, gray] ($ (boxtop)+(0.5*\b,0) $) arc[start angle=0, end angle=-10, radius=0.5*\b] node[midway, left] (angle) {};
            \draw[thin, gray, shorten >=-0.2cm] (angle.center) .. controls ++(-0.2,0.02) and ++(0.2,-0.0) .. ++(-0.3,0.4) node [left]{$\varphi$};
        \end{scope}
        \begin{scope}[shift={(-1.5,-4.5)}]
            \draw[ultra thick] (-30:2cm) arc (-30:30:2cm) node[near end](temp_for_radius) {};
            \draw[loosely dashed, thin, gray] ($ (210:2cm) + (3.95,0) $) arc (210:150:2cm) node [midway](contact_old) {} ;
            \draw[ultra thick] ($ (210:2cm) + (3.95,-0.5) $) arc (210:150:2cm) node[midway](contact_new) {};
            \draw[Stealth-] (temp_for_radius.center) -- ($ (temp_for_radius.center)!.5!(0,0) $) node[midway, above left] {$R$};
            \draw[Stealth-Stealth] ([xshift=0.15cm]contact_old.center) -- ([xshift=0.15cm]contact_new.center) node[midway, right] {$\Delta{}d$};
        \end{scope}
    \end{tikzpicture}%
        \caption{Bending deformation.}
        \label{fig:analytical_model_bending}     
    \end{subfigure}
    \caption{Analytical model to describe the influence of the particle size for shear and bending deformation.}
    \label{fig:analytical_model}
\end{figure}

The numerical results of Section~\ref{sec:results:particle_size} suggest that the observed damping behavior is (approximately) independent of the particle size. To confirm these numerical observations by analytical arguments, first-order models for different deformation modes of particle-filled cavities will be derived in this section. To allow for an analytical treatment of the problem, the following simplifying assumptions are made: Since high packing densities are present inside the cavity, it is assumed that the particles are always in contact. With the particles always in contact, the model only considers dissipation from friction and not from particle impacts. Furthermore, no dissipation between particles and walls is considered (which holds true, e.g., if the particles are sticking to the walls). In the following, a pure shear mode and a pure bending mode are considered. Thereto, assume a rectangular box filled with equal sized particles of radius $R$ and normal forces at the walls $F_{N,x}$ and $F_{N,y}$  (see Fig.~\ref{fig:analytical_model_undeformed}), which are a direct consequence of the powder compaction process. The normal forces at each contact point $F_{N,x,i}$ and $F_{N,y,i}$ yield
\begin{equation}
    F_{N,y,i} = \frac{F_{N,y}}{N_x} \quad \text{and} \quad F_{N,x,i} = \frac{F_{N,x}}{N_y},
\end{equation}
with the number of particles in x- and y-direction $N_x = b/(2R)$, $N_y = l/(2R)$.

A linear shear deformation with angle $\varphi$ (see Fig.~\ref{fig:analytical_model_shear}) results in a relative motion 
\begin{equation}
    \Delta{}d = 2R \cdot \tan(\varphi)
\end{equation}
between two contacting particles. The work contribution $W_{x,i}$ for one particle is given by
\begin{equation}
    W_{x,i} = F_{T,x,i} \Delta{}d \quad \text{with} \quad F_{T,x,i} = \mu F_{N,y,i} ,
\end{equation}
where $F_{T,x,i}$ is the friction force acting in x-direction and $\mu$ the coefficient of friction. Assuming that all contacts are subject to sliding friction, the work is dissipated. Then, the total dissipation is the sum over the work of all contacts $i$ that have relative motion
\begin{equation}
    \begin{split}
        W &= \sum_i W_{x,i} = \sum_i  \mu \frac{F_{N,y}}{N_x} \Delta{}d = N_x (N_y - 1) \mu \frac{F_{N,y}}{N_x} \Delta{}d \\
        &\approx N_y \mu F_{N,y} \Delta{}d = \frac{l}{2R} \mu F_{N,y} 2R \tan(\varphi) = l \mu F_{N,y} \tan(\varphi),
    \end{split}
\end{equation}
using that there are $N_x = l/(2R)$ contacts in x-direction and $N_y -1 \approx N_y = b/(2R)$ contacts in y-direction. This results means that the dissipation for the shear case is independent of the particle radius. Halving the particle size increases the number of contact points by a factor of four, but also halves the contact force and the relative sliding distance such that the total work remains the same within this model.

For pure bending, a linear deformation is considered (see Fig.~\ref{fig:analytical_model_bending}). The deformation is idealized such that all particles remain in contact. There is no displacement in x-direction and the displacement in y-direction is given by
\begin{equation}
    \Delta{}y = - \Delta{}y_{max} \frac{x}{b/2}\frac{y}{l/2} = f \cdot x y,
\end{equation}
where $\Delta{}y_{max} = b/2 \cdot \tan(\varphi)$ is the maximum displacement at the corners and $f = 4\Delta{}y_{max} / (bl)$ is introduced to summarize the constant factors. The relative motion between two particles (with coordinates $(x_1,y)$ and $(x_2,y)$ in the undeformed configuration) is then given by
\begin{equation}
    \Delta{}d = f x_2 y - f x_1 y = f (x_2 - x_1) y = f \cdot 2R \cdot y,
\end{equation}
where $x_2 - x_1 = 2R$ such that the relative motion between the particles depends on their y-position. Note that there is only relative motion between contacting particles which are at the same height in the undeformed configuration, i.e., at the same y-coordinate. Again, the work contribution $W_{x,i}$ for one pair of particles is calculated
\begin{equation}
    W_{y,i} = \Delta{}d_i F_{T,y,i} \quad \text{with} \quad F_{T,y,i} = \mu F_{N,x,i}, 
\end{equation}
with the friction force $F_{T,y,i}$ in y-direction. Again, it is assumed that all particles remain in contact and are in sliding friction. The total dissipation results in
\begin{equation} \label{eq:bending:work:i}
    W = \sum_i W_{y,i} = \sum_i \Delta{}d_i F_{T,y,i} = \sum_i f \cdot 2R \cdot y_i \cdot \mu F_{N,x,i} = f \cdot 2R \cdot \mu \frac{F_{N,x}}{N_y} \sum_i  y_i .
\end{equation}
The sum can be further simplified. There are $N_x - 1$ contacts in x-direction with the same $y$ value such that the remaining sum is only over $N_y$ contacts. Here, symmetry with respect to the x-axis can be used and the y-coordinate is expressed in multiples of the particle diameter $y_i = 2R\cdot i$. For the resulting sum the summation $\sum_{i=1}^{n} i = \frac{n(n+1)}{2}$ can be used. This results in
\begin{equation} \label{eq:bending:work:sum}
        \sum_i  y_i = (N_x - 1) \sum_{i_y} y_{i_y} = (N_x - 1) \sum_{i_y=1}^{N_y/2} 2R \cdot i_y = (N_x - 1)2R \frac{\frac{N_y}{2}(\frac{N_y}{2} + 1)}{2} \approx 2R N_x \frac{N_y^2}{8} ,
\end{equation}
where $N_x - 1 \approx N_x$ and $\frac{N_y}{2}(\frac{N_y}{2} + 1)\approx \frac{N_x}{4}$ for large $N$.
Finally, by combining~\eqref{eq:bending:work:i} and~\eqref{eq:bending:work:sum} the total work results in
\begin{equation}
    W = f \cdot 2R \cdot \mu \frac{F_{N,x}}{N_y} \cdot 2R N_x \frac{N_y^2}{8} = f \cdot \mu F_{N,x} \cdot 4R^2 \cdot \frac{1}{8} \frac{b}{2R} \frac{l}{2R} = \frac{f b l}{8} \cdot \mu F_{N,x} .
\end{equation}
Again, the result is independent of the particle size. Of course, the real particle behavior is more complex. The particle size follows a size distribution. Shear and bending modes occur at the same time and are not linear. However, the results show that the influence of the particle size is small compared to the effect of the packing density (which directly influences the normal forces $F_{N,x}$ and $F_{N,y}$). So, for the manufacturing of particle dampers, getting the right packing density inside the cavity is more important than choosing the particle size.

\section{Experimental realization}
This section presents a first proof of concept regarding an experimental realization of the particle damper systems considered in the numerical studies. In particular, it shall be demonstrated that the time scale of damping (,i.e., the oscillation time to standstill), which is the relevant characteristic for most practical applications, is in the same order of magnitude as observed in the numerical studies. For these first-order comparisons, it is sufficient to take DEM model parameters (e.g., contact stiffness, coefficient of friction, etc.) from the literature instead of performing time-consuming model calibration experiments~\cite{Meier2019}.

\begin{figure}[htb]
    \centering
    \begin{subfigure}[t]{0.45\textwidth}
        \begin{tikzpicture}
            \newcommand{\annotation}[4]{\node[align=left, fill=white, fill opacity=0.8, text opacity=1, font=\small] (#2) at (#3) {#1};\draw[-stealth, thick, red] (#2) -- (#4);}
            \begin{scope}[xshift=1.5cm]
                \node[anchor=south west,inner sep=0] (image) at (0,0) {\includegraphics[width=\textwidth]{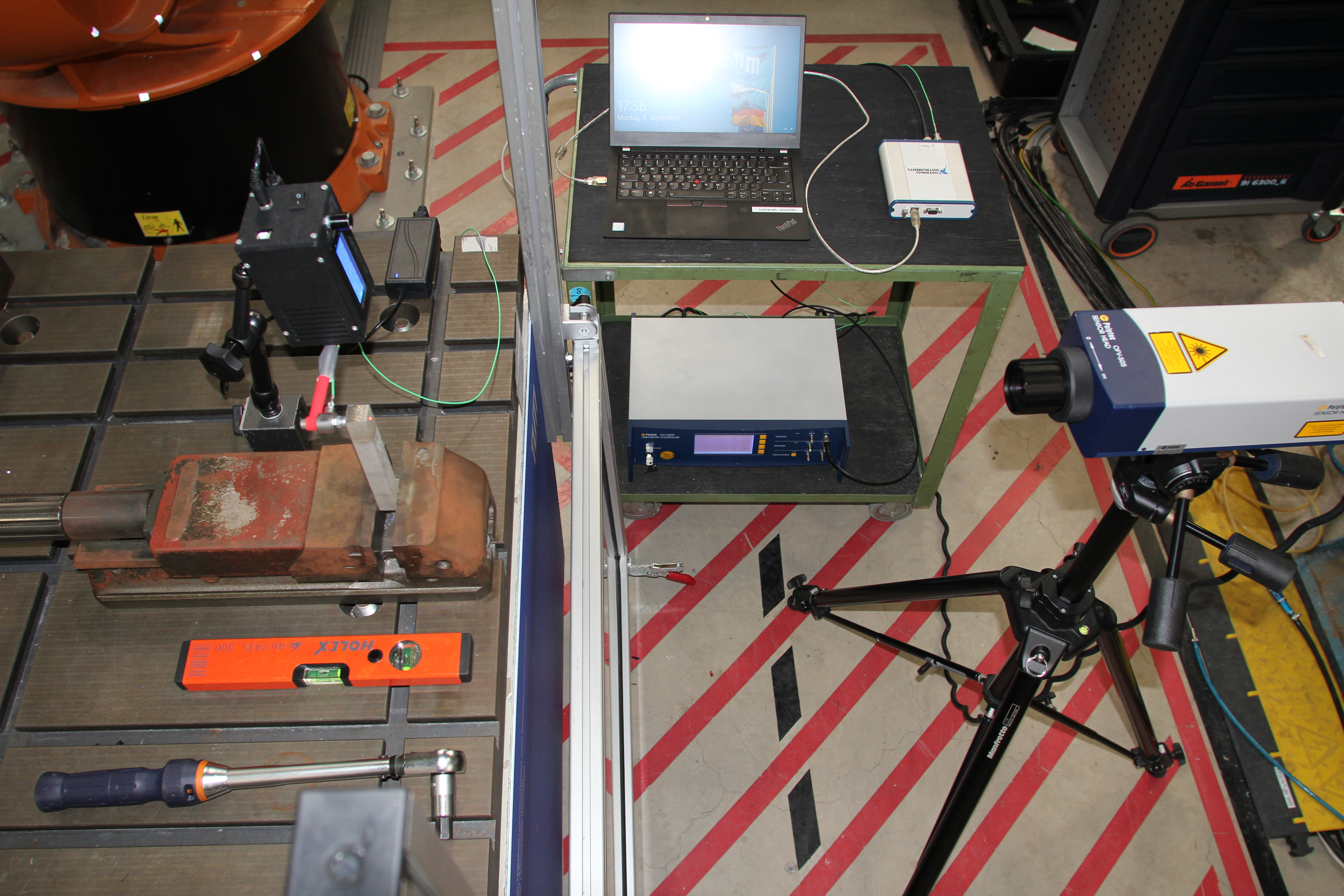}};
                \begin{scope}[x={(image.south east)},y={(image.north west)}]
                    \annotation{Laser vibrometer}{vibrometer}{0.8,0.2}{0.9,0.55}
                    \annotation{Specimen}{specimen}{0.4,0.15}{0.28,0.48}
                    \annotation{Automatic \\ impulse hammer}{hammer}{0.17,0.88}{0.24,0.68}
                    \annotation{Force sensor}{sensor}{0.13,0.17}{0.25,0.52}
                    \annotation{Measurement\\system}{measure}{0.85,0.88}{0.53,0.8}
                \end{scope}
            \end{scope}
        \end{tikzpicture}%
        \caption{Experimental setup.}
        \label{fig:experimental_setup}
    \end{subfigure}
    \begin{subfigure}[t]{0.45\textwidth}
        \includegraphics[width=\textwidth]{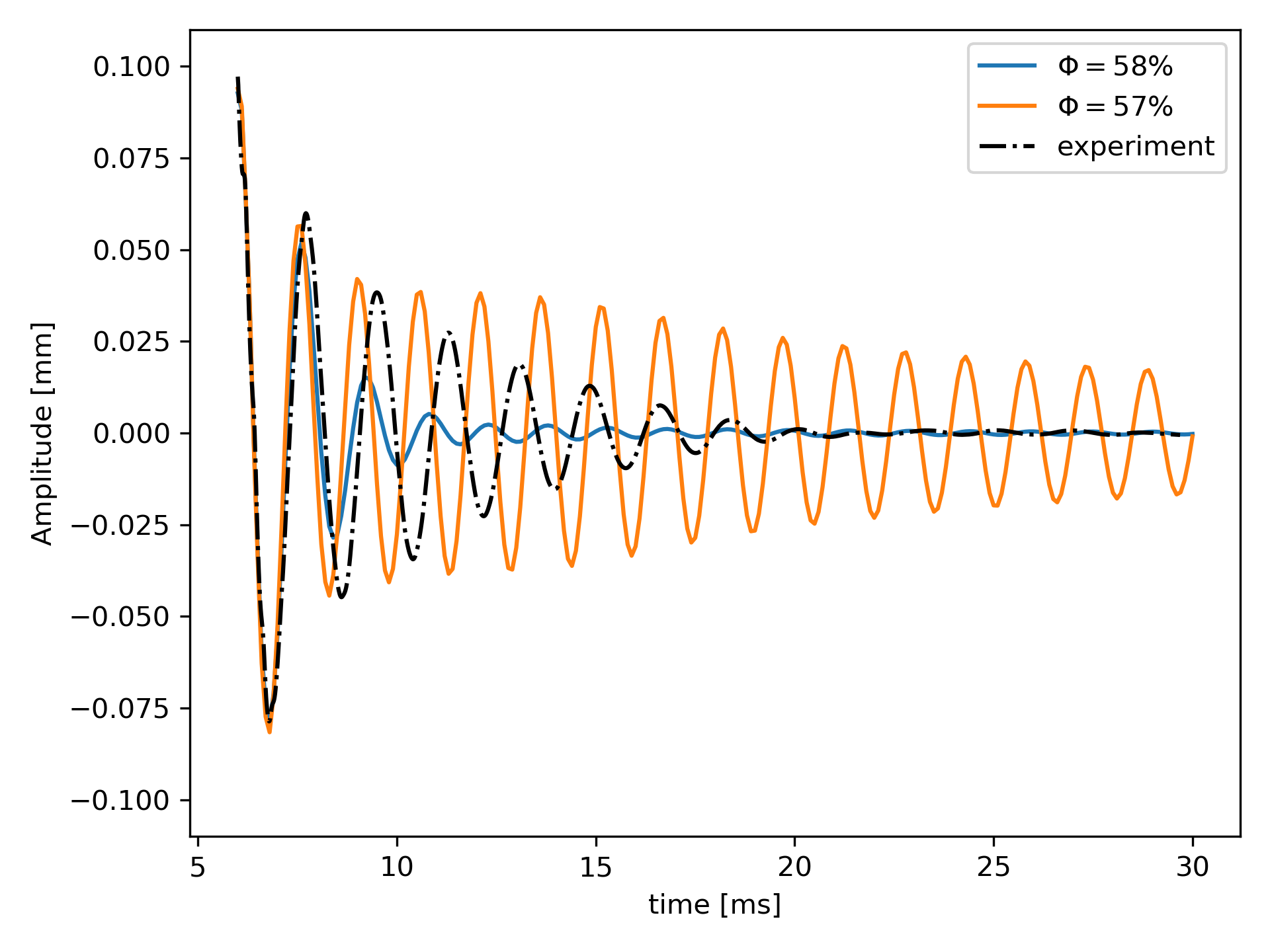}
        \caption{Comparison of experiment and simulation for the excitation of a standing beam.}
        \label{fig:exp_results}
    \end{subfigure}
    \caption{Experimental realization.}
    \label{fig:exp_realization}
\end{figure}

Figure~\ref{fig:experimental_setup} shows the experimental setup. Beams with dimensions $20 \times 20 \times 150~\si{\milli\metre\cubed}$ and a closed cavity of $18 \times 18 \times 110~\si{\milli\metre\cubed}$ at the center are printed out of stainless steel 316L. This geometry resembles the one used for the numerical study. The first~\SI{20}{\milli\meter} of the beam are clamped in a vice such that the beam is oriented vertically. The beam is excited on one side by an automatic impulse hammer with an attached force sensor (Dytran  1051 V4). On the opposite side a laser vibrometer (Polytec Sensor Head OFV-505 and Polytec Vibrometer Controller OFV-5000) measures the velocity near the top of the beam. The use of an automatic impulse hammer allows high repeatability of the excitation.

In a first study, the beam is excited with a force of~\SI{108}{\newton}. The measured displacement curve is compared to the corresponding simulation results. Given that the DEM model parameters have not been calibrated to match the experimental powder material, the simulated damping behavior agrees well with the experimental measurements. In particular, the time scale of damping (i.e., time to standstill), which is relevant for many practical applications, shows a very good agreement. The experimental result lies between the simulation results with a packing density of $\Phi = \SI{57}{\percent}$ and $\Phi = \SI{58}{\percent}$ (both for the particle size $D50 = \SI{1.0}{\milli\meter}$) and is of the same order of magnitude. For a future, quantitative comparison, relevant model parameters, e.g., contact stiffness, coefficient of friction, and material damping, need to be calibrated with experiments. Moreover, the packing density in experiments needs to be measured with high fidelity.

\section{Conclusion and Outlook}
\label{sec:conlusion}
In the present work, a two-way coupled discrete element - finite element model has been proposed to study the fundamental dissipation mechanisms in additively manufactured particle dampers. In particular, the proposed DEM-FEM framework allowed for the first time to consistently describe the interaction between oscillating deformable structures and enclosed powder packings, revealing for the considered system that sliding friction between powder particles, as imposed by the deformable cavity walls, is the main mechanism of dissipation. Simulations of the free oscillation of hollow cantilever beams with enclosed powder packings showed that there is an optimal packing density at which the best damping is achieved. For the powder sizes studied in this work, the optimal packing density ranged from~\SI{58}{\percent} to~\SI{61}{\percent} for the unfused powder. Packing densities different from the optimal packing densities yielded significantly worse damping. This strong dependence of the particle damper with the packing density should be considered when additively manufacturing particle dampers. Further, the results showed only a small influence of different particle sizes on the dissipation which could be verified by first-order models of the shear and bending mode of a powder cavity. Cohesive forces between particles were found to have no noticeable influence on the damping. Similarly, the influence of the coefficient of restitution, i.e., the parameter defining the dissipation from impacts, was small. In turn, the coefficient of friction had a large influence where the best damping was achieved at a coefficient of friction $\mu=0.4$. A first comparison of the simulated damping behavior agreed well with experiments.

Future work will be concerned with the calibration of the model with experimental data. In addition, the goal is to optimize the design of the powder cavities in terms of size, shape, and positioning within the part in order to achieve desired damping properties of components.

\section*{Acknowledgements}
The authors acknowledge funding of this work by the Deutsche Forschungsgemeinschaft (DFG, German Research Foundation) within project 414180263.

\appendix

\section{Computational Modeling Parameters}
\label{app:parameters}
See Table~\ref{tab:parameters_dem} and~\ref{tab:parameters_fem}.

\begin{table}[hbt]
    \centering
    \caption{Tabulated values for the parameters of the DEM model}
    \label{tab:parameters_dem}
    \begin{tabular}{llll}
        \toprule
        Symbol & Parameter & Value & Unit \\
        \midrule
        $\rho$ & Density & 8000 & $\si{\kilo\gram\per\meter\cubed}$\\
        $k_N$ & Penalty parameter & 5642 & \si{\newton\per\meter} \\
        $c_g$ & Maximum relative penetration & 0.055 & $-$ \\
        $\mu$ & Coefficient of friction & 0.4 & $-$ \\
        $e$ & Coefficient of restitution & 0.4 & $-$ \\
        $\nu$ & Poisson's ratio & 0.27 & $-$ \\
        $g$ & Gravity & 9.81 & \si{\meter\per\second\squared} \\
        $\Delta{}t$ & Time step size & \num{2.5e-6} & \si{\second} \\
        \midrule
        Original Log-normal particle size distribution \\
        $D_{50}$ & Median & 0.0265646 & \si{\milli\meter} \\
        $\sigma$ & Sigma & 0.2707 & $-$ \\
        $D_{90}$ & $90^{th}$ percentile & 0.0375828 & \si{\milli\meter} \\
        $D_{10}$ & $10^{th}$ percentile & 0.0187766 & \si{\milli\meter} \\
        Scaled Log-normal particle size distribution \\
        $D_{50}$ & Median & 1.0 & \si{\milli\meter} \\
        $\sigma$ & Sigma & 0.2707 & $-$ \\
        $r_{max}$ & Maximum cutoff radius & 0.707385 & \si{\milli\meter} \\
        $r_{min}$ & Minimum cutoff radius & 0.353414 & \si{\milli\meter} \\
        \bottomrule        
    \end{tabular}
\end{table}

\begin{table}[hbt]
    \centering
    \caption{Tabulated values for the FEM model}
    \label{tab:parameters_fem}
    \begin{tabular}{llll}
        \toprule
        Symbol & Parameter & Value & Unit \\
        \midrule
        $\rho$ & Density & 8000 & \si{\kilo\gram\per\meter\cubed} \\
        $E$ & Young's modulus & 175 & \si{\giga\pascal} \\
        $\nu$ & Poisson's ratio & 0.27 & $-$ \\
        $h$ & Element size (hex8) & \num{1.0} & \si{\milli\meter} \\
        $\epsilon$ & Tolerance DEM-FEM coupling & \num{1.0e-3} & $-$ \\
        \midrule
        Algorithmic parameters of Generalized-$\alpha$ method \\
        $\rho_{\infty}$ & Spectral radius & $0.8$ & $-$ \\
        $\alpha_f$ &  & $0.\overline{4}$ & $-$ \\
        $\alpha_m$ &  & $0.\overline{3}$ & $-$ \\
        $\beta$ &  & 0.30864 & $-$ \\
        $\gamma$ &  & 0.61111 & $-$ \\
        \bottomrule        
    \end{tabular}
\end{table}

\bibliographystyle{elsarticle-num} 
\bibliography{references}

\end{document}